\documentclass[final,5p,times,twocolumn,authoryear]{elsarticle}

\usepackage{amsmath}
\usepackage{amssymb}

\DeclareMathOperator{\tr}{tr}
\allowdisplaybreaks

\usepackage{url}

\usepackage[acronym,nomain,nogroupskip,nonumberlist]{glossaries}
\makeglossaries
\newacronym{awph}{AWPH}{Astrophysical Weighted Particle Hydrodynamics}
\newacronym{sph}{SPH}{Smoothed Particle Hydrodynamics}
\newacronym{wpm}{WPM}{Weighted Particle Methods}
\newacronym{rmd}{RMD}{Renormalised Meshless Derivative}
\newacronym{mls}{MLS}{Moving Least Squares}
\newacronym[shortplural=KDEs]{kde}{KDE}{Kernel Density Estimation}
\newacronym{mci}{MCI}{Monte Carlo Integration}
\newacronym[shortplural=PDFs]{pdf}{PDF}{Probability Density Function}
\newacronym{dk}{DK}{Dimensionless Kernel}
\newacronym{mfv}{MFV}{Meshless Finite Volume}
\newacronym{mfm}{MFM}{Meshless Finite Mass}

\newcommand{\mymat}[1]{\mathrm{\mathbf{#1}}}
\newcommand{\myvec}[1]{\mathbf{#1}}
\newcommand{\Alpha}{\mathrm{A}}
\newcommand{\Beta}{\mathrm{B}}
\newcommand{\ndim}{\nu}
\newcommand{\idmat}{\mymat{I}}

\journal{Astronomy $\&$ Computing}

\begin{document}

\begin{frontmatter}



\title{An improvement to the spatial discretisation in the Astrophysical Weighted Particle Hydrodynamics method}


\author[first,second]{Alex Massaro Achá}
\ead{aacha759@gmail.com}
\author[first,second]{Isaac Alonso Asensio}
\author[first,second]{Claudio Dalla Vecchia}
\affiliation[first]{organization={Instituto de Astrofísica de Canarias},
            addressline={C/Vía Láctea s/n},
            city={La Laguna},
            postcode={E-38205},
            state={Tenerife},
            country={Spain}}
\affiliation[second]{organization={Departamento de Astrofísica, Universidad de La Laguna},
            addressline={Av. Astrofísico Francisco Sánchez s/n},
            city={La Laguna},
            postcode={E-38206},
            state={Tenerife},
            country={Spain}}

\begin{abstract}
The discretisation of space in the \acrlong{awph} method is attained through a partition-of-unity-based cubature rule. The node (particle) weights in such rules must in general be obtained using approximations, for which only one has been thus far proposed in the literature. In this work, new approximations for these weights are proposed, their properties are studied, and their impact on the accuracy of the cubature rule and the overall method is assessed. Approximations for node centroids are also proposed and studied, and their possible applications are laid out. Additionally, for schemes in which the kernel-function compact-support size is set to be constant, rather than adaptive, weight and centroid approximations are provided as well. Finally, for the cubature rule arising from the weight approximation previously proposed in the literature, an asymptotic expression for its mean relative error is derived, and its usefulness for understanding the finite-sample properties of the cubature rule is showcased with some simple examples.
\end{abstract}




\end{frontmatter}




\section{Introduction}
\label{sec:introduction}
The \gls{awph} method of \citet{gn11} is a meshless numerical scheme conceived to solve the equations of inviscid compressible hydrodynamics and magnetohydrodynamics for applications in Astrophysics. It is based on the scheme developed in \citet{v99}, which is itself built from the \gls{sph} method \citep[e.g.][and references therein]{s10b}.

\citet{v99} coined the term \gls{wpm} to refer to meshless numerical methods for conservation laws that share the \gls{sph} strategy of using a weighted sum over nodes (i.e., particles, points, tracers, etc.) to approximate the sought-off solution. It then introduced a variant of \gls{wpm} which aimed at solving two of the major drawbacks of \gls{sph}: its shock-capturing properties and its estimation of derivatives. To address the former, the weak form of the conservation law is used as a starting point for the discretisation; while for the latter, a so-called \gls{rmd} is introduced which computes derivatives of affine functions exactly. This \gls{wpm} variant was later called the weak \gls{rmd} scheme in subsequent studies \citep{m99,l01}, and had its stability, consistency and convergence studied in \citet{lv08a,lv08b}. It is this variant that was used in \citet{gn11} to construct the \gls{awph} method.

Common to all \gls{wpm}-like methods is the discretisation of the spatial integral appearing in the weak form of a conservation law through the introduction of a \textit{cubature} rule (i.e., a quadrature rule for a multivariate integral). Hence, at the beginning of a calculation, a weight needs to be assigned to each node such that the resulting cubature rule approximates the spatial integral of the initial state of a problem to some desired degree of accuracy. For the subsequent time evolution, then, it is necessary to prescribe how nodes and weights will evolve, hopefully in such a way that the newly obtained cubature rule maintains its accuracy in approximating the spatial integral of the time-evolved state. \citet{v99} advocated to move the nodes with the local fluid velocity and to update their weights with an advection equation (e.g., its Equation 2.2). \citet{gn11}, on the other hand, while still moving the nodes with the local fluid velocity, resorted to a partition-of-unity approach to obtain their weights.

In Applied Mathematics, partitions of unity are well known in the field of interpolation. For example, they appear naturally in the construction of a \gls{mls} approximation of a function. Moreover, since interpolation techniques form the basis of meshless numerical methods for solving partial differential equations, partitions of unity have also made their appearance in the latter \citep[e.g.][]{b96}. In the context of numerical integration, the use of partitions of unity to obtain cubature rules seems to have been pioneered by \citet{c02} and \citet{d02}. Its properties for the integration of convex functions were later formalised by \citet{g06}.

Due to computational-efficiency considerations, it is common for partitions of unity to be built from functions having a compact support. Moreover, in some applications it is useful to make the size of this compact support adaptive. Such is the case for the \gls{awph} method. A criterion needs to be defined, then, that prescribes the way in which the compact-support size will adapt, which should be controlled by one or more numerical parameters.

Then, by choosing a particular partition-of-unity function and, if desired, a scheme for the adaptivity of its compact-support size, a cubature rule is completely established and its weights can be calculated given the node positions. An approximation for them for the choices in \gls{awph} was put forward by \citet{gn11}, which was thenceforth adopted in all the subsequent work on the method \citep{h15,h18,a23,g23,k24}.

The role played by the cubature weights, however, is not limited to the spatial discretisation. They are, in fact, part of the definition of the \gls{rmd} \citep[e.g.][Equation 14]{lv08a}, through which they play a role in gradient estimation and in establishing the interaction strength between two neighbouring nodes \citep[e.g.][Equation 23]{lv08a}. Furthermore, in fluid-flow calculations, they are directly used to obtain mass-density estimates, which are in turn used to obtain the fluid's pressure field. As such, they play a role in the dynamics of the flow as well.

In this work, an approach is put forward which allows obtaining higher-order corrections to the weight approximation proposed by \citet{gn11}. Several new approximations are obtained by truncating the expressions at different orders, which then have their accuracy evaluated by comparing them to a computationally expensive exact weight calculation. The approach is also used to obtain several new approximations for node centroids, which are also evaluated following the same method as for the weights. After choosing the most suitable weight approximations arising from the analysis, their impact on the accuracy and convergence of \gls{awph}['s] cubature rule is evaluated for some simple cases of numerical integration, and their impact on the full solution of a conservation law is showcased with a specific example.

Additionally, the partition-of-unity-based cubature rule arising from the weight approximation proposed by \citet{gn11} is shown to be equivalent to a Monte Carlo integration scheme, which enables obtaining an asymptotic expression for its mean relative error. An example of the usefulness of this expression to understand the finite-sample properties of such a cubature rule is then given.

Finally, for applications in which the kernel's compact-support size is set to be a constant, rather than adaptive, a different approach is used to derive new weight and centroid approximations, which are given in \ref{ap:consth}.

It should be noted that the treatment of boundary conditions is completely disregarded in this work. As such, Cartesian periodic boundary conditions are assumed throughout.

The paper is structured as follows. Section \ref{sec:awph} introduces the relevant details of the \gls{awph} method for this work. Section \ref{sec:kdemci} establishes the equivalence between \gls{awph}['s] cubature rule, with the weights proposed by \citet{gn11}, and Monte Carlo integration. Section \ref{sec:approxwc} outlines the derivation of the new weight and centroid approximations that are the core of this work. Section \ref{sec:propwc} addresses the accuracy of these weight and centroid approximations. Section \ref{sec:propcb} provides a simple numerical assessment of the accuracy and convergence of \gls{awph}['s] cubature rule. Section \ref{sec:sedov} showcases the effects of the weight approximations in the solution of a problem in inviscid compressible hydrodynamics. Finally, in Section \ref{sec:discussion} the results are discussed, and Section \ref{sec:conclusions} lays out the conclusions.

\section{The spatial discretisation in \gls{awph}}
\label{sec:awph}
The weak form of a general scalar conservation law can be written as
\begin{equation}
    \int \left[ u(\myvec{x},t) \dot{\varphi}(\myvec{x},t) + F^{\alpha}(u) \varphi_{,\alpha}(\myvec{x},t) - S(u) \varphi(\myvec{x},t) \right] d\myvec{x}\,dt = 0,
    \label{eq:weak}
\end{equation}
\noindent where $u(\myvec{x},t)$ is the unknown function, $F^{\alpha}(u)$ are the components of the flux vector, $S(u)$ collects the possible source terms, and $\varphi(\myvec{x},t)$ is an arbitrary test function. The notation $\dot{\varphi}$ means a total time derivative along a trajectory $\myvec{x}(t)$, $\varphi_{,\alpha}$ means a partial derivative with respect to the $\alpha$ spatial coordinate, and summation is implied in the term with repeated Greek indices. The integration is done over the entire spatial and temporal domain of the problem. Thus, the first step in \gls{wpm}, and by extension in \gls{awph}, is to discretise this spatial integral using a cubature rule.

A cubature rule is defined as a set of point-weight pairs $\left\{(\myvec{x}_i,w_i)\right\} $ such that
\begin{equation}
    \int_\Omega f(\myvec{x},t) \, d\myvec{x} \approx \sum_{i=1}^{N} w_i(t) \, f(\myvec{x}_i,t)
    \label{eq:cubature}
\end{equation}
\noindent to some desired degree of accuracy. More general cubature rules can include point-weight pairs associated to values of the function's derivatives, besides the ones associated to values of the function itself. However, such rules are not relevant for the present work.

Equation \ref{eq:cubature} is typically required to be exact for functions $f(\myvec{x},t)$ belonging to some function space that captures the general properties of the functions that the rule will be applied to in practice. The most general one is the space of polynomials of a given degree. All in all, the most basic requirement is for the rule to be exact for constant functions, which implies that $\sum_i w_i = \int d\myvec{x} = |\Omega|$; i.e., the sum of the weights must equal the size of the domain. The assumption of exactness for polynomials up to a given degree is ubiquitous in mathematical proofs of the accuracy and convergence of quadrature and cubature rules \citep[e.g.][]{e80,d84}.

Partitions of unity are sets of functions $\left\{\psi_i(\myvec{x})\right\}$ that satisfy the following \citep[][Definition 3.3]{g06}
\begin{enumerate}[i.]
    \item $\psi_i(\myvec{x}) \ge 0$,
    \item $\int \psi_i(\myvec{x}) \, d\myvec{x} > 0$, and
    \item $\sum_i \psi_i(\myvec{x}) = 1 $.
\end{enumerate}
\noindent Their usefulness in the present context lies in that, given a set of points $\left\{\myvec{x}_i\right\} $, they can be used to divide a given domain $\Omega$ into a set of overlapping subdomains $ \left\{\Omega_i\right\} $ such that
\begin{enumerate}[i.]
    \item $\left|\Omega_i\right| > 0$,
    \item $\bigcup_i \Omega_i=\Omega$,
\end{enumerate}
\noindent simply by setting
\begin{equation}
    \left|\Omega_i\right| = w_i = \int_\Omega \psi_i(\myvec{x}) \, d\myvec{x},
    \label{eq:puweight}
\end{equation}
\noindent since it immediately follows that $\sum_i w_i = \int_\Omega d\myvec{x} = \left|\Omega\right|$. Hence, the combined set $\left\{(\myvec{x}_i,\psi_i(\myvec{x}))\right\} $ defines a cubature rule for numerical integration in $\Omega$ that is, at worst, exact for constant functions. Furthermore, it is shown in \citet{g06}\footnote{The full proof consists of Proposition 2.1, Definition 2.2, (see also Remark 2.3), and finally Theorems 3.4 and 3.8.} that if the set $\left\{(\myvec{x}_i,\psi_i(\myvec{x}))\right\} $ has the additional property of
\begin{equation}
    \myvec{x}_i = \myvec{c}_i = \dfrac{\int_\Omega \myvec{x} \, \psi_i(\myvec{x}) \, d\myvec{x}}{\int_\Omega \psi_i(\myvec{x}) \, d\myvec{x}},
    \label{eq:pucentroid}
\end{equation}
\noindent then the cubature rule it defines will be exact for affine functions as well. The vector $\myvec{c}_i$ will here be called the \textit{centroid} of node $\myvec{x}_i$. It is also commonly called the centre of mass or barycentre.

In practice, however, Equations \ref{eq:puweight} and \ref{eq:pucentroid} must typically be approximated, either analytically or numerically. As a consequence, the integration of constant and affine functions will not be exact; instead, they will inherit the accuracy of the approximations being used.

In \gls{awph}, the functions composing the partition of unity are given as
\begin{equation}
    \psi_i(\myvec{x},H) = \dfrac{W(\left|\myvec{x} - \myvec{x}_i\right|, H)}{\omega(\myvec{x},H)},
    \label{eq:awphpu}
\end{equation}
\noindent where $W(x,H)$ is called the kernel and weight function in the \gls{sph} and the \gls{mls} literatures, respectively, $H$ is the size of its compact support, and
\begin{equation}
    \omega(\myvec{x},H) = \sum_j W(|\myvec{x} - \myvec{x}_j|, H).
    \label{eq:omega}
\end{equation}
\noindent This choice for a partition of unity corresponds to the shape function of an \gls{mls} function approximation that is exact for constant functions. They are also called Shepard functions due to the work of \citet{s68}. The spatial adaptivity of the compact-support size $H$ is given through the following constraint
\begin{equation}
    C_\ndim \, H^\ndim \, \omega(\myvec{x},H) = N_\mathrm{eff},
    \label{eq:awphh}
\end{equation}
\noindent where $\ndim$ is the space dimensionality and $C_\ndim = 1, \,\pi, \,4\pi/3$ for $\ndim = 1,\,2,\,3$. Equation \ref{eq:awphh} implicitly defines the function $H(\myvec{x})$, whose values must in general be obtained by an iterative procedure, and $N_\mathrm{eff}$ is the numerical parameter controlling its adaptivity. For monotonically decreasing kernel functions, $H(\myvec{x})$ is a monotonically increasing function of $N_\mathrm{eff}$.

Having prescribed $H(\myvec{x})$, weights can in principle be evaluated by plugging Equation \ref{eq:awphpu} into Equation \ref{eq:puweight}. However, as noted earlier, this can only be done in an approximate way. \citet{gn11} proposed the approximation
\begin{equation}
    w_i \approx \dfrac{1}{\omega(\myvec{x}_i,H(\myvec{x}_i))}.
    \label{eq:awphweight}
\end{equation}

Hereafter, for the sake of brevity, the kernel function will be referred to simply as the \textit{kernel}, and its compact-support size simply as the \textit{support}. Furthermore, the $H(\myvec{x})$ function will be referred to as the \textit{support function}.

\section{Relation with kernel density estimation and Monte Carlo integration}
\label{sec:kdemci}
As it turns out, the cubature rule that follows from the relations described thus far can be cast into the formalisms of \gls{kde} and \gls{mci}. This brings considerable insight into its properties and provides a theoretical basis for interpreting the results obtained with it in the simple numerical-integration examples of Section \ref{sec:propcb}.

It can be seen that the quantity $\omega(\myvec{x},H(\myvec{x}))$ in Equation \ref{eq:omega} is proportional to an adaptive \gls{kde} of the underlying \gls{pdf} of node positions. Evidently, denoting by $\hat{g}(\myvec{x},H(\myvec{x}))$ the adaptive \gls{kde} of the node-position \gls{pdf} $g(\myvec{x})$,
\begin{equation}
    \hat{g}(\myvec{x},H(\myvec{x}))=\dfrac{\omega(\myvec{x},H(\myvec{x}))}{N},
    \label{eq:akde}
\end{equation}
\noindent where $N$ is the number of nodes. Furthermore, Equation \ref{eq:awphh} has its origin in the so-called nearest-neighbour method \citep[][Chapter 5]{s86} used as a so-called \textit{balloon} estimator \citep{ts92} in adaptive \gls{kde}, in which the support (commonly called the \textit{band width} in \gls{kde} literature) is taken as a function of position, $H(\myvec{x})$, in order to define its adaptivity. As such, the $N_\mathrm{eff}$ parameter is related, approximately and statistically, to the number of nodes $i$ fulfilling $|\myvec{x} - \myvec{x}_i| \le H(\myvec{x})$; that is, to the \textit{number of neighbours} at position $\myvec{x}$.

The previous observations lead to the conclusion that the cubature rule, with weights computed using Equation \ref{eq:awphweight}, is actually an importance-sampling \gls{mci} scheme in which the importance-sampling distribution is set to an adaptive \gls{kde} of the node-position \gls{pdf}. Indeed,
\begin{equation}
    \int f(\myvec{x}) \, d\myvec{x} \approx \sum_{i=1}^N \dfrac{f(\myvec{x}_i)}{\omega(\myvec{x}_i,H(\myvec{x}_i))} = \dfrac{1}{N} \sum_{i=1}^N \dfrac{f(\myvec{x}_i)}{\hat{g}(\myvec{x}_i,H(\myvec{x}_i))}.
    \label{eq:ismci}
\end{equation}
\noindent As a consequence, it inherits all the properties of such schemes. Most importantly, it will be the most accurate when the importance-sampling distribution $\hat{g}(\myvec{x},H(\myvec{x}))$ is as close as possible to a \gls{pdf} given by \citep[e.g.][Theorem 6.5]{e00}
\begin{equation}
    g(\myvec{x}) = \dfrac{|f(\myvec{x})|}{\int |f(\myvec{z})| \, d\myvec{z}},
\end{equation}
\noindent which implies that node positions should look like samples drawn from the latter as much as possible. Furthermore, even in the ideal case where this condition is fulfilled, the rule's accuracy will be limited by that with which $\hat{g}(\myvec{x},H(\myvec{x}))$ approximates $g(\myvec{x})$. In fact, if such a favourable case is indeed assumed, then results from \gls{kde} theory can be used to obtain an asymptotic approximation of the mean relative error of the cubature rule. The details are given in \ref{ap:relerr}. The final result is the following
\begin{equation}
    \mathrm{MRE} \approx \Theta \dfrac{N_\mathrm{eff}^{2/3}}{N^{2/3}} + \Phi \dfrac{N_\mathrm{eff}^{4/3}}{N^{4/3}} + \dfrac{\Xi}{N_\mathrm{eff}},
    \label{eq:meanrelerr}
\end{equation}
\noindent where $\Theta$, $\Phi$ and $\Xi$ are constants that depend on the kernel and the function being integrated. The terms in the right-hand side come from the bias, squared bias and variance of the \gls{kde}, respectively. The first observation to make is that, asymptotically, the mean relative error is not a monotonic function of the $N_\mathrm{eff}$ parameter. This is a direct consequence of the bias-variance trade-off as manifested in the \gls{kde}. The second observation is that, whenever the variance term dominates, no accuracy improvement can be expected by increasing $N$ while keeping $N_\mathrm{eff}$ constant, which is to say that the cubature rule will show no convergence in such cases.

It should be noted, however, that while Equation \ref{eq:meanrelerr} provides useful insight into the asymptotic (in the sense detailed in \ref{ap:relerr}) properties of the cubature rule, it should be used with care when interpreting such properties under finite samples, as the asymptotic assumption is formally never fulfilled. In Section \ref{sec:propcb}, for example, it is shown that it can be successfully used as a regression model to understand the observed relative errors of the cubature rule while numerically integrating some simple functions. It should be stressed as well that this result is only valid for cubature weights computed as proposed by \citet{gn11} (that is, Equation \ref{eq:awphweight}). It does not hold for the approximations derived in this work, as the weights then cease to be proportional to a \gls{kde} of the node-position \gls{pdf}.

\section{Weight and centroid approximations}
\label{sec:approxwc}

\subsection{Derivation}
\label{sec:derivation}
In deriving approximations for weights and centroids, it proves convenient to introduce what will here be called the \gls{dk} as
\begin{equation}
    \zeta(|\myvec{x}-\myvec{x}_i|,H(\myvec{x})) = \dfrac{W(|\myvec{x}-\myvec{x}_i|,H(\myvec{x}))}{H^\ndim(\myvec{x})}.
    \label{eq:dk}
\end{equation}
\noindent Furthermore, it will be assumed that the \gls{dk} can always be written as
\begin{equation}
    \zeta(|\myvec{x}-\myvec{x}_i|,H(\myvec{x})) = \zeta\left(\dfrac{|\myvec{x}-\myvec{x}_i|}{H(\myvec{x})}\right) = \zeta(q_i(\myvec{x})).
    \label{eq:dkqi}
\end{equation}
\noindent Then, in terms of it, Equation \ref{eq:awphpu} can be rewritten, with the help of Equation \ref{eq:awphh}, as
\begin{equation}
    \psi_i(\myvec{x},H(\myvec{x})) = C_\ndim \, N_\mathrm{eff}^{-1} \, \zeta(|\myvec{x}-\myvec{x}_i|,H(\myvec{x})).
    \label{eq:dkpu}
\end{equation}

Additionally, for the case of centroids, instead of attempting to evaluate Equation \ref{eq:pucentroid} directly, it proves simpler to evaluate the vector
\begin{equation}
    \myvec{s}_i = \int_\Omega (\myvec{x} - \myvec{x}_i) \, \psi_i(\myvec{x},H(\myvec{x})) \, d\myvec{x},
    \label{eq:svec}
\end{equation}
\noindent from which, then, $\myvec{c}_i = \myvec{s}_i/w_i + \myvec{x}_i$.

With these simplifications, the approach followed goes as follows.
\begin{enumerate}[i.]
    \item Expand the support function in a Taylor series around a node position $\myvec{x}_i$, discarding third- and higher-order terms.
    \item Expand $\zeta(|\myvec{x}-\myvec{x}_i|,H(\myvec{x}))$ in a Taylor series around the node's support $H(\myvec{x}_i)$, discarding fourth- and higher-order terms.
    \item Substitute the support-function expansion into the $\zeta(|\myvec{x}-\myvec{x}_i|,H(\myvec{x}))$ expansion. This yields an approximation for the latter where the support function has been completely replaced by $H(\myvec{x}_i)$ and its first and second spatial derivatives evaluated at $\myvec{x}_i$.
    \item Evaluate Equations \ref{eq:puweight} and \ref{eq:svec} using Equation \ref{eq:dkpu} and introducing the $\zeta(|\myvec{x}-\myvec{x}_i|,H(\myvec{x}))$ approximation of the previous point. As it turns out, this can be done analytically.
\end{enumerate}

The final results can be cast in a concise way as follows
\begin{gather}
    w_i \approx \dfrac{1}{\omega(\myvec{x}_i,H_i)}\left( 1 + \alpha_\zeta \Alpha_i + \beta_\zeta \Beta_i + \gamma_\zeta \Gamma_i\right),
    \label{eq:impawphweight} \\
    \myvec{s}_i \approx \dfrac{1}{\omega(\myvec{x}_i,H_i)} \left( \alpha_\zeta \mymat{A}_i + \beta_\zeta \mymat{B}_i + \gamma_\zeta \mymat{\Gamma}_i \right) \myvec{g}_i
    \label{eq:awphcentroid}
\end{gather}
\noindent where lower- and upper-case bold-face letters denote column vectors and matrices, respectively, and $\myvec{g}_i = \myvec{g}(\myvec{x}_i)$ is the support-function's gradient vector. The explicit forms of each of the coefficients are given in \ref{ap:coeff} for the one-, two- and three-dimensional cases.

\subsection{Spatial derivatives of the support function}
\label{sec:hderivatives}
In order to be able to use the above-derived approximations, the support-function's gradient vector, $\myvec{g}$, and Hessian matrix, $\mymat{H}$, need to be known. They can be derived analytically by differentiating Equation \ref{eq:awphh}. It proves convenient to rewrite said equation using the \gls{dk} as follows
\begin{equation}
    \sum_j \zeta(|\myvec{x}-\myvec{x}_j|,H(\myvec{x})) = C_\ndim^{-1} \, N_\mathrm{eff}.
    \label{eq:dkawphh}
\end{equation}
\noindent Then, by computing the latter's total differential twice, the following are obtained
\begin{gather}
    \myvec{g}(\myvec{x}) = -\dfrac{\boldsymbol{a}(\myvec{x})}{\mathcal{N}(\myvec{x})},
    \label{eq:gradh} \\
    \mymat{H}(\myvec{x}) = -\dfrac{\boldsymbol{\mathcal{A}}(\myvec{x}) + \boldsymbol{\mathcal{B}}(\myvec{x}) + \boldsymbol{\mathcal{C}}(\myvec{x})}{\mathcal{N}(\myvec{x})},
    \label{eq:hessh}
\end{gather}
\noindent where
\begin{align}
    \boldsymbol{a}^\alpha(\myvec{x}) &= \sum_j \zeta_{,\alpha}(q_j), \label{eq:calA} \\
    \mathcal{N}(\myvec{x}) &= \sum_j \zeta_{,H}(q_j), \\
    \boldsymbol{\mathcal{A}}^{\alpha\beta}(\myvec{x}) &= \sum_j \zeta_{,\alpha\beta}(q_j), \\
    \boldsymbol{\mathcal{B}}^{\alpha\beta}(\myvec{x}) &= \myvec{g}^\alpha(\myvec{x}) \sum_j \zeta_{,\beta H}(q_j) + \myvec{g}^\beta(\myvec{x}) \sum_j \zeta_{,\alpha H}(q_j), \\
    \boldsymbol{\mathcal{C}}^{\alpha\beta}(\myvec{x}) &= 
    \myvec{g}^\alpha(\myvec{x}) \, \myvec{g}^\beta(\myvec{x}) \sum_j \zeta_{,HH}(q_j). \label{eq:calC}
\end{align}
\noindent in which superscripts denote specific entries in vectors and matrices, and
\begin{align}
    &\zeta_{,\alpha}(q_i) = \dfrac{\partial \zeta(q_i)}{\partial x^\alpha} = \dfrac{1}{H(\myvec{x})}\dfrac{x^\alpha - x_i^\alpha}{|\myvec{x}-\myvec{x}_i|} \zeta_{,q}(q_i), \\
    &\zeta_{,H}(q_i) = \dfrac{\partial \zeta(q_i)}{\partial H} = -\dfrac{|\myvec{x}-\myvec{x}_i|}{H^2(\myvec{x})} \zeta_{,q}(q_i), \\
    &\zeta_{,\alpha\beta}(q_i) 
    \begin{aligned}[t]
        &= \dfrac{\partial^2 \zeta(q_i)}{\partial x^\beta \partial x^\alpha} = \dfrac{1}{H^2(\myvec{x})} \left\{ \dfrac{H(\myvec{x})}{|\myvec{x}-\myvec{x}_i|} \zeta_{,q}(q_i) \, \delta^{\alpha\beta} \right. \\
        &+ \left. \dfrac{(x^\alpha - x_i^\alpha) (x^\beta - x_i^\beta)}{|\myvec{x}-\myvec{x}_i|^2} \! \left[ \zeta_{,qq}(q_i) - \dfrac{H(\myvec{x})}{|\myvec{x}-\myvec{x}_i|} \zeta_{,q}(q_i) \right] \right\},
    \end{aligned} \\
    &\zeta_{,\alpha H}(q_i) = \dfrac{\partial^2 \zeta(q_i)}{\partial H \partial x^\alpha} = -\dfrac{x^\alpha - x_i^\alpha}{H^3(\myvec{x})} \! \left[ \zeta_{,qq}(q_i) \! + \! \dfrac{H(\myvec{x})}{|\myvec{x}-\myvec{x}_i|} \! \zeta_{,q}(q_i) \! \right], \\
    &\zeta_{,HH}(q_i) = \dfrac{\partial^2 \zeta(q_i)}{\partial H^2} = \dfrac{|\myvec{x}-\myvec{x}_i|^2}{H^4(\myvec{x})} \! \left[ \zeta_{,qq}(q_i) \! + \! 2\dfrac{H(\myvec{x})}{|\myvec{x}-\myvec{x}_i|} \zeta_{,q}(q_i) \right],
\end{align}
\noindent in which $\zeta_{,q}(q_i) = d\zeta(q_i)/dq$, and likewise for $\zeta_{,qq}(q_i)$ (see Equation \ref{eq:dkqi} for the definition of $q$).

It is important to note that when evaluating Equations \ref{eq:calA} to \ref{eq:calC} at node positions, $\myvec{x}=\myvec{x}_i$, there will be a term in their right-hand-side summations for $j=i$. Hence, these equations are meaningful provided that their limit when $\myvec{x} \to \myvec{x}_i$ exist. In order for this to be the case, the limit must exist for their $j=i$ term. This can be seen to hold, as
\begin{align}
    &\lim_{\myvec{x} \to \myvec{x}_i} \zeta_{,\alpha}(q_i) = 0, \\
    &\lim_{\myvec{x} \to \myvec{x}_i} \zeta_{,H}(q_i) = 0, \\
    &\lim_{\myvec{x} \to \myvec{x}_i} \zeta_{,\alpha\beta}(q_i) = \dfrac{\zeta_{,qq}(0)}{H_i^2} \, \delta^{\alpha\beta}, \label{eq:dkwtf}\\
    &\lim_{\myvec{x} \to \myvec{x}_i} \zeta_{,\alpha H}(q_i) = 0, \\
    &\lim_{\myvec{x} \to \myvec{x}_i} \zeta_{,HH}(q_i) = 0.
\end{align}
\noindent It should be noted that the validity of Equation \ref{eq:dkwtf} is actually a requirement the kernel must fulfil in order for the Hessian matrix to be defined at node positions\footnote{The actual requirement is that $\lim_{q\to0} \zeta_{,q}/q = \zeta_{,qq}(0)$, of which Equation \ref{eq:dkwtf} is a direct consequence.}. Accordingly, it has been verified to hold for all the kernels considered in this work that are used to compute a Hessian matrix.

\section{Properties of the weight and centroid approximations}
\label{sec:propwc}
Having derived new approximations for weights and centroids, this section is devoted to studying their accuracy. Noting that, in fact, several approximations have been derived, as the Taylor-series expansions that were used to obtain them can be truncated at different orders, the goal is to assess the merit of each of these truncation orders in terms of the accuracy and computational cost and complexity of the weight and centroid approximations they lead to, with the expectation of finding the most suitable.

\subsection{The setup}
The procedure followed consists in comparing the different weight and centroid approximations to their corresponding ``exact'' values. These exact values are taken as those obtained from the evaluation of the integrals in Equations \ref{eq:puweight} and \ref{eq:pucentroid} with an arbitrarily accurate cubature rule. For the latter, the Cuhre routine of the \textsc{Cuba} library\footnote{\url{https://feynarts.de/cuba}} is used, which is described in \citet{h05} and references therein. This same routine is also used to evaluate Equations \ref{eq:puweight} and \ref{eq:pucentroid} for the case in which only the \gls{dk} is expanded in a Taylor series, while the support function is used in full. This gives an upper bound on the improvement that can be robustly achieved by Equations \ref{eq:impawphweight} and \ref{eq:awphcentroid}, as further improvement must come from error compensation between the two Taylor-series expansions, which would be coincidental and thus not numerically robust. Furthermore, given the high computational cost associated with evaluating integrals with such cubature rules, this process is carried out in two spatial dimensions.

In order to proceed, a labelling scheme must be defined for the relevant truncation orders. The notation $\partial^n \zeta \ \partial^m H$ is used where $n$ and $m$ denote the orders of the \gls{dk} and support-function Taylor-series expansions, respectively. Thus, for example, $\partial^3\zeta \ \partial^\infty H$ denotes the weight and centroid approximations resulting from expanding the \gls{dk} to third order and using the support function in full (this approximation was mentioned in the previous paragraph), while $\partial^1\zeta \ \partial^1 H$ denotes the approximations resulting from first-order expansions for both the \gls{dk} and the support function. It should be mentioned that, as a consequence of the \gls{dk} and its derivatives being even functions of $\myvec{x}-\myvec{x}_i$, half of the contributions to Equations \ref{eq:impawphweight} and \ref{eq:awphcentroid} from the combination of terms in the Taylor series' vanish identically. This is why not all possible truncation-order combinations are shown. For instance, the weight approximations $\partial^0\zeta \ \partial^0 H$ and $\partial^1\zeta \ \partial^1 H$ are equivalent, since the integration of the additional term that is present the latter vanishes identically. This is the approximation previously proposed in the literature (Equation \ref{eq:awphweight}). Finally, in the fully expanded forms of Equations \ref{eq:impawphweight} and \ref{eq:awphcentroid}, terms corresponding to a given order of the \gls{dk} expansion can be easily identified, as those that have a combined power in the gradient and Hessian-matrix components that is linear, quadratic or cubic must come from the first-, second- or third-order terms in the \gls{dk} expansion, respectively.

Discrepancies between approximations and exact values are quantified through measures of dispersion and central tendency (that is, bias). These are taken as the mean of the unsigned and signed errors, respectively, where the latter are defined as
\begin{equation}
    \sigma_i = \left| e_i \right|, \qquad \quad e_i = \dfrac{\hat{x}_i-x_i}{y_i},
\end{equation}
\noindent where $x$ denotes either the weights or centroid components, $\hat{x}$ being an approximation and $x$ the exact value, and $y$ is a suitable scale that is chosen to ease interpretation. These choices are always indicated explicitly in all the figures of this section.

\subsection{Node distributions and numbers}
\label{sec:nodedist}
Node positions are randomly sampled from three different distributions. In all cases, each coordinate is sampled independently. The base-line case is a uniform distribution. Then, in an attempt to mimic the conditions found in real-world problems, a piecewise-uniform and a normal distribution are considered. The piecewise-uniform distribution is taken as
\begin{equation}
    p(x) = 
    \begin{cases}
        1/2 & \mathrm{if} \ 0 < x < 1/2, \\
        3/2 & \mathrm{if} \ 1/2 < x < 1, \\
        0   & \mathrm{otherwise},
    \end{cases}
\end{equation}
\noindent which is used to sample one of the coordinates, while the other is sampled from the uniform distribution. This intends to mimic discontinuities, which are ubiquitous in non-linear conservation laws. Finally, the normal distribution has $\sigma = 0.1$ and a mean located at the centre of the domain. This distribution, in turn, intends to mimic node clustering. Node numbers $N$ are $200^2$ for the uniform and piecewise-uniform distributions, while the normal distribution has $300^2$.

\subsection{The kernel functions}
\label{sec:thekernels}
In general, the degree to which the support function can be approximated by its Taylor series will improve if its derivatives can be made smaller in magnitude. The latter, as a consequence of the support-function's definition through Equation \ref{eq:awphh}, depend not only on the node distribution, but on the magnitude of the kernel derivatives as well. Thus, it is to be expected that kernels with smaller slopes and curvatures (that is, less centrally concentrated) will yield better approximations, provided they meet the requirements to be Taylor expanded up to the relevant orders.

To demonstrate the above, the following kernels, here ordered by increasing central concentration, are considered. The $W_{3/2}^H$ kernel of \citet{c08}, which is written here in \gls{dk} form as \citep[see][Equation 3]{c08}
\begin{equation}
    W_{3/2}^H(q) = B_\nu
    \begin{cases}
        1 & \mathrm{if} \ q = 0, \\
        \left( \dfrac{\sin \pi q}{\pi q} \right)^{3/2} & \mathrm{if} \ 0 < q \le 1, \\
        0 & \mathrm{otherwise},
    \end{cases}
    \label{eq:sincdk}
\end{equation}
\noindent where $B_\nu$ is the normalisation constant for $\nu$ dimensions, which must be obtained numerically. Additionally, the $\phi_{1,0}$ kernel of \citet[][Table 1]{wu95}, and the standard cubic- and quartic-spline kernels \citep[e.g.][Table 1]{da12} are considered.

With the exception of the $W_{3/2}^H$ kernel, they all have continuous second derivatives, and thus can be used for the approximations at all the relevant truncation orders. The second derivative of the $W_{3/2}^H$ kernel, however, diverges at $q = 1$, thus it can at most be used for the $\partial^2\zeta \ \partial^1 H$ approximations.

\subsection{Shortlisting the truncation orders}
Of all the possible combinations of an up-to third-order \gls{dk} expansion and an up-to second-order support-function expansion, several do not merit further investigation given the undesired properties or poor accuracy exhibited by their weight and/or centroid approximations.

In the case of the weight approximation, it is seen that any approximation that makes use of the diagonal entries of the support-function's Hessian matrix has the tendency to yield substantial overestimates. Although the overestimation is reduced by increasing the $N_\mathrm{eff}$ parameter, these approximations remain far from being competitive with those that do not include these entries. This property is a consequence of the fact that the support function can become strongly concave up inside local clusters of nodes, in which case its second-order Taylor series grossly overestimates it near the boundaries of the region delimited by a node's support, thus leading to the weight overestimation. Nonetheless, it turns out that approximations in which the diagonal entries of the Hessian matrix are discarded, but the off-diagonal ones are kept, are still competitive. Thus, hereafter, all the weight approximations that are second order in the support function discard the diagonal entries of the Hessian matrix, and keep only the off-diagonal ones. They are denoted by appending the suffix ``$\mathrm{(nd)}$'' (meaning \textit{no diagonal}) to their labels.

Furthermore, still in the case of weight approximations, in most applications weights are required to be positive, as their most common physical interpretation is the size (that is, length, area or volume) of a node. This becomes a difficulty for the $\partial^3 \zeta \ \partial^2 H\mathrm{(nd)}$ approximation, since it can indeed yield negative weights. Thus, in this work the choice was made to address this problem by introducing a minimum value that this approximation can take, and setting this minimum to the value given by the $\partial^1 \zeta \ \partial^1 H$ approximation. A different approximation could indeed be used in place of the latter; however, given that the $\partial^1 \zeta \ \partial^1 H$ approximation is on average a clear underestimate of the true weight (as per the results of this section), it is at the very least a sensible choice for the purpose of limiting the $\partial^3 \zeta \ \partial^2 H\mathrm{(nd)}$ approximation. The resulting approximation is denoted as $\max\left( \partial^3 \zeta \ \partial^2 H \mathrm{(nd)}, \ \partial^1 \zeta \ \partial^1 H \right)$.

On the other hand, in the case of the centroid approximation, it should first be noted that obtaining a centroid approximation requires a weight approximation to be obtained beforehand (see Equations \ref{eq:pucentroid} and \ref{eq:svec}). In principle, this gives the freedom to choose different truncation orders for the $w_i$ and $\myvec{s}_i$ approximations from which the centroid $\myvec{c}_i$ is obtained. However, in this work this freedom is not explored, and thus the same truncation orders are always used to obtain $w_i$ and $\myvec{s}_i$. That being the case, it turns out that centroid approximations that include the third-order term of the \gls{dk}['s] Taylor series (that is, $\partial^3 \zeta \ \partial^1 H$ and $\partial^3 \zeta \ \partial^2 H$) yield poor accuracy when compared to approximations that exclude this term, and thus they will not be considered further. Likewise, even though it performs better that the previous two, the $\partial^1 \zeta \ \partial^1 H$ approximation still does not compete with the rest, and thus it is excluded as well.

\subsection{Results}

\begin{figure*}[t]
    \centering
    \includegraphics[width=\textwidth,keepaspectratio]{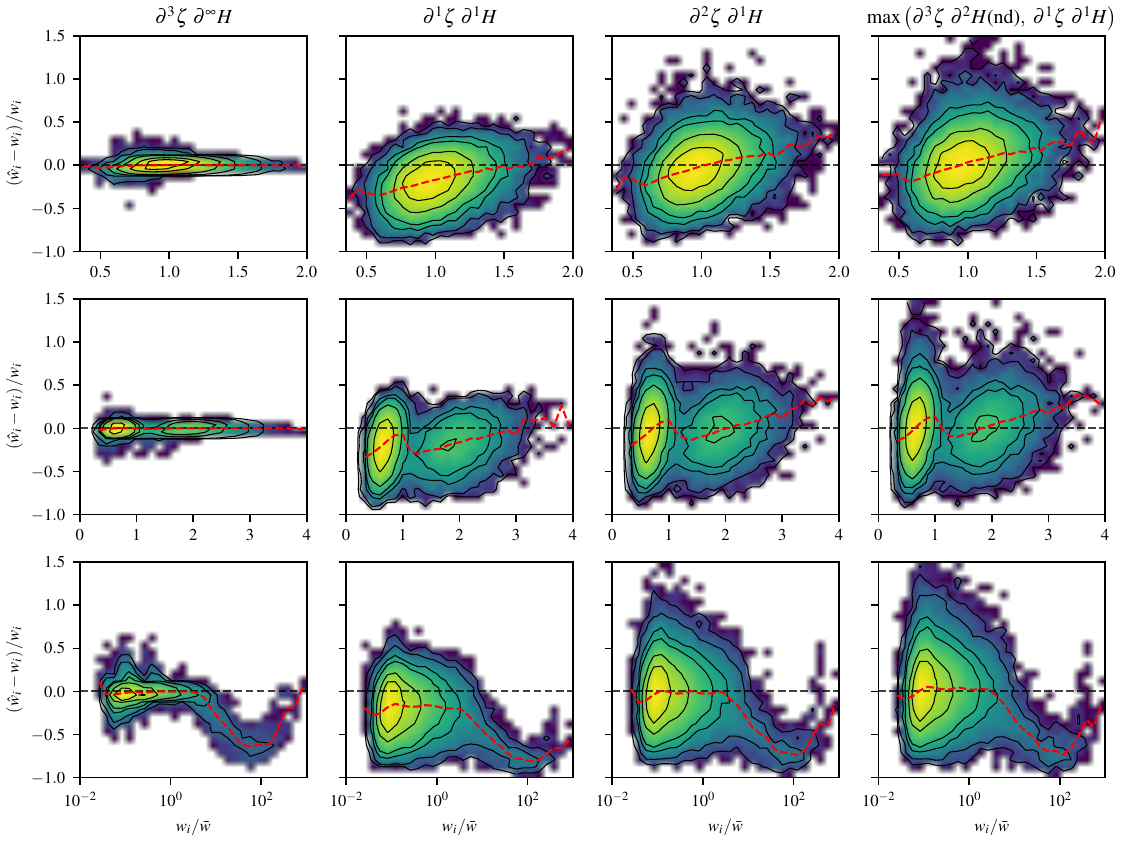}
    \caption{Weight-approximation relative-error distributions. Nodes are sampled from the uniform (top row), piecewise-uniform (middle row) and normal (bottom row) distributions, as described in Section \ref{sec:nodedist}. Columns correspond to different weight approximations, as indicated at the top of the Figure. Contours enclose roughly 40, 70, 90, 95, 99 and 99.7\% of the total number of points. The horizontal axis is normalised by $\bar{w}$, which denotes the mean weight for each distribution. Note that the horizontal axes have different ranges in each row. The dashed black line marks the locus of vanishing error, while the dashed red line is the median in horizontal-axis bins. These results correspond to the $\phi_{1,0}$ kernel of \citet{wu95} with $N_\mathrm{eff} = 20$.}
    \label{fig:weights}
\end{figure*}

\begin{figure*}[t]
    \centering
    \includegraphics[width=\textwidth,keepaspectratio]{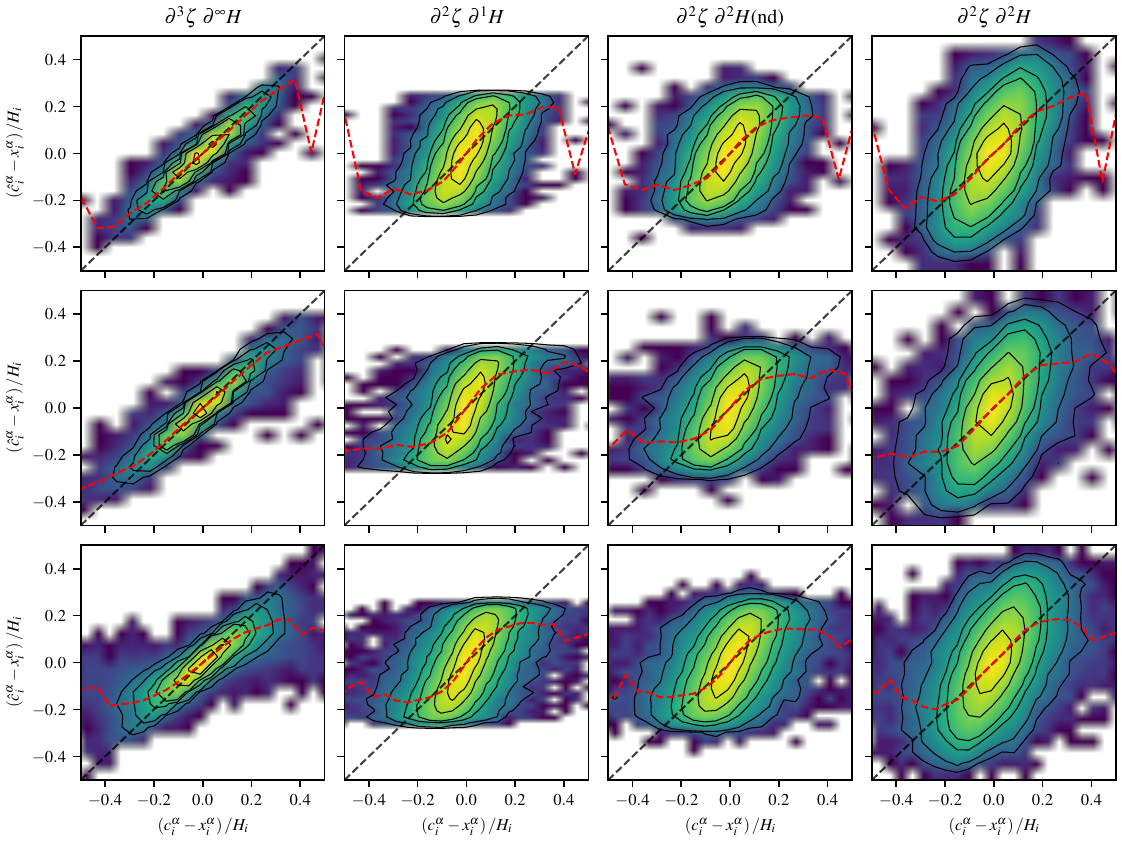}
    \caption{Joint distributions of centroid approximations and exact values, both in units of node supports. Nodes are sampled from the uniform (top row), piecewise-uniform (middle row) and normal (bottom row) distributions. Columns correspond to different centroid approximations, as indicated at the top of the Figure. Contours enclose roughly 40, 70, 90, 95, 99 and 99.7\% of the total number of points. Centroid components are lumped together in the uniform- and normal-distribution cases, while only the component across the discontinuity is included in the piecewise-uniform case. The dashed black line marks the locus of vanishing error, while the dashed red line is the median in horizontal-axis bins. These results correspond to the $\phi_{1,0}$ kernel of \citet{wu95} with $N_\mathrm{eff} = 20$.}
    \label{fig:centroids}
\end{figure*}

\begin{figure*}
    \centering
    \includegraphics[width=\textwidth,keepaspectratio]{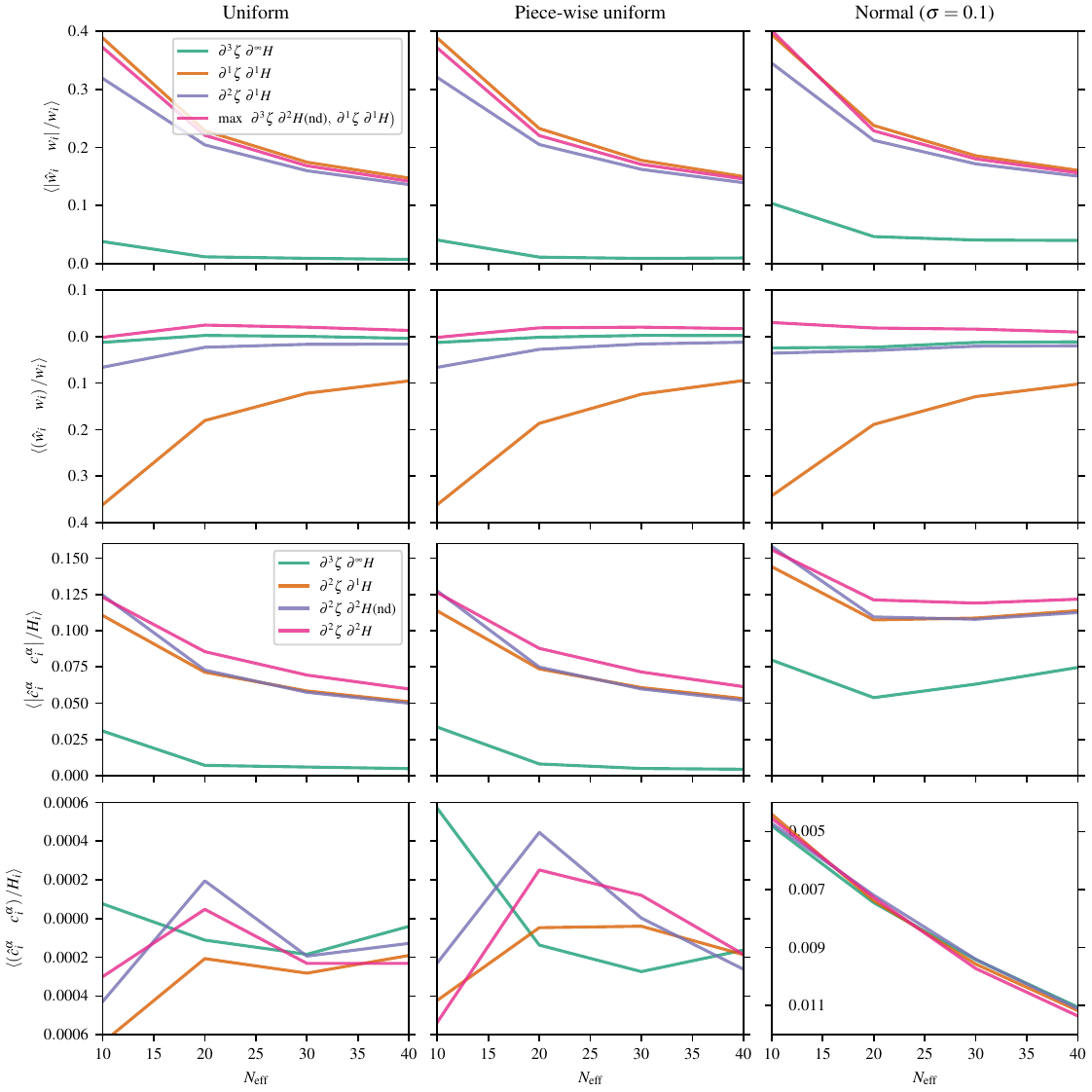}
    \caption{Error dispersion and central tendency (bias), as indicated by the labels in the vertical axes, in the weight and centroid approximations as a function of $N_\mathrm{eff}$. Columns correspond to different node distributions as described in Section \ref{sec:nodedist}, which are indicated at the top of the Figure. Note the different vertical-axis range in the bottom-right panel. These results correspond to the $\phi_{1,0}$ kernel of \citet{wu95}.}
    \label{fig:scores_wu10}
\end{figure*}

\begin{figure*}[t]
    \centering
    \includegraphics[]{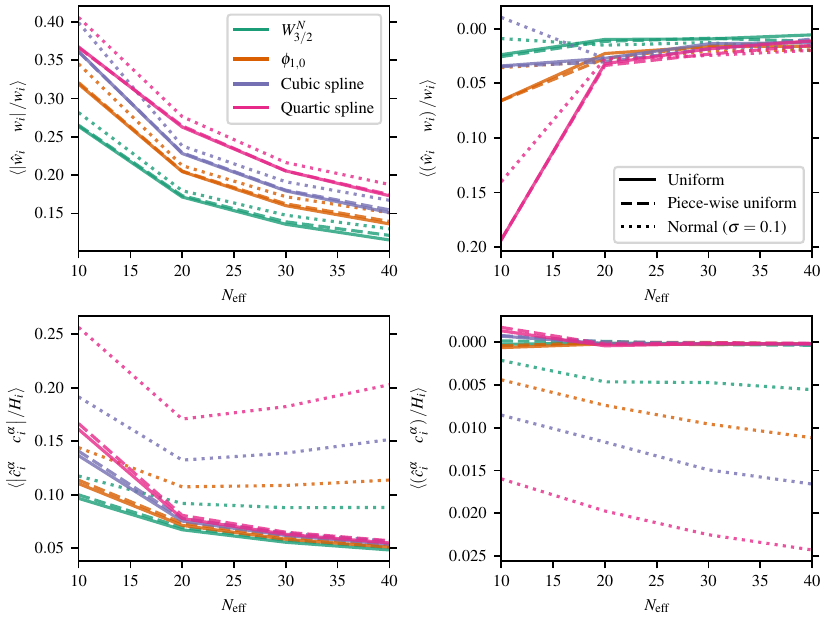}
    \caption{Error dispersion and central tendency (bias), as indicated by the labels in the vertical axes, in the $\partial^2 \zeta \ \partial^1 H$ weight and centroid approximations, as a function of $N_\mathrm{eff}$. Results are shown for the kernels mentioned in Section \ref{sec:thekernels} and the node distributions described in Section \ref{sec:nodedist}, as indicated in the legends.}
    \label{fig:scores_all}
\end{figure*}

Figures \ref{fig:weights} and \ref{fig:centroids} show the distributions of the shortlisted weight and centroid approximations with respect to the exact values for the three node distributions described in Section \ref{sec:nodedist}. In the case of weights, Figure \ref{fig:weights} shows the distributions of relative errors as a function of the exact weights. In the case of centroids, Figure \ref{fig:centroids} shows the joint distributions of approximations and exact values, both normalised by node supports.

The first thing to notice in Figure \ref{fig:weights} is that the $\partial^1 \zeta \ \partial^1 H$ approximation (the one previously proposed in the literature) has a bias towards underestimating the weights in all three node distributions being considered, and that the $\partial^2 \zeta \ \partial^1 H$ and $\max\left( \partial^3 \zeta \ \partial^2 H \mathrm{(nd)}, \ \partial^1 \zeta \ \partial^1 H \right)$ approximations have this bias reduced. Furthermore, perhaps unsurprisingly, in all cases the $\partial^3 \zeta \ \partial^\infty H$ approximation performs much better both in terms of bias and dispersion with respect to the exact values. However, the difference is less stark in the case of normally distributed nodes, particularly for the nodes with the largest weights, where all approximations yield substantial underestimates of the exact weights. These correspond to the nodes sampled at the outskirts of the normal distribution, and the reason behind such poor performance can be understood as follows. The support-function definition guarantees that every point $\myvec{x}$ in the domain has at least one node inside its support $H(\myvec{x})$. Thus, every point in the domain will contribute to the weight and centroid integrals of at least one node. However, when the \gls{dk} is Taylor expanded with respect to the support, the \gls{dk} itself and all its derivatives are evaluated only at the support value around which the expansion is made; the latter being in practice the node supports $H_i$. Due to the compactly supported property of the \gls{dk}[s] that are used in practice, this means that points for which $\left| \myvec{x} - \myvec{x}_i \right| > H_i$ for all $i$ will fail to contribute to any of the weight and centroid integrals. This effect should be most noticeable in nodes around regions of low node density, where the support function has local maxima. In the case of the normal distribution considered here, this effect is exacerbated since a considerable fraction of the domain is completely devoid of nodes (the value of $\sigma = 0.1$ was purposely chosen in order to illustrate this effect). Naturally, increasing the order of the Taylor expansion should eventually compensate for this effect. However, for the third-order expansion considered here, it still leads to underestimates of both weights and centroid components as well, as discussed next.

In Figure \ref{fig:centroids}, the first thing to notice is that all approximations are somewhat biased towards underestimating the centroids when these are far away from node positions, which is to say that, in such cases and on average, they tend to yield centroids closer to node positions than they truly are. This is especially noticeable in the normal distribution, where nodes sampled at its outskirts have centroids that are considerably far away from their positions\footnote{These are not shown in Figure \ref{fig:centroids} for clarity reasons. In the normally distributed case, the horizontal axis actually extends roughly to $[-10,10]$, while the vertical-axis extent is as shown. However, the fraction of nodes in these excluded regions, as the contour lines indicate, is roughly 0.3\%.}, for which all approximations yield substantial underestimates. The reason behind this is the same as for the weight underestimation in this same distribution, which was described in the previous paragraph. In contrast, in nodes whose centroids are closer to their positions, the approximations, albeit with a fair amount of dispersion, are much less biased and, on average, closely follow the one-to-one line. Finally, it should be mentioned that the $\partial^2 \zeta \ \partial^1 H$ approximation is actually bounded in the following sense
\begin{equation}
    \left| \dfrac{\hat{c}_i^\alpha - x_i^\alpha}{H_i} \right| \le \sqrt{\dfrac{\alpha_\zeta}{\nu+1}},
\end{equation}
\noindent where $\alpha_\zeta$ is the constant appearing in Equations \ref{eq:impawphweight} and \ref{eq:awphcentroid}, for which values are given in \ref{ap:coeff}, and $\nu$ is the number of dimensions. In the case of Figure \ref{fig:centroids}, in which the $\phi_{1,0}$ kernel of \citet{wu95} is used in two dimensions, the right-hand side evaluates to roughly $0.25$.

A summary of the error properties of the shortlisted approximations as a function of the $N_\mathrm{eff}$ parameter is shown in Figure \ref{fig:scores_wu10}, for the case of the $\phi_{1,0}$ kernel of \citet{wu95}. Looking at the first and third rows, from top to bottom, it is seen that the dispersions of the approximations around the exact values, for both weight and centroids, are comparable (excluding $\partial^3 \zeta \ \partial^\infty H$, of course), and that $\partial^2 \zeta \ \partial^1 H$ is slightly favoured, again for both weight and centroids. Furthermore, looking at the bottom row, it is seen that all centroid approximations have similar small overall biases. In fact, such small values are likely affected by the accuracy of the cubature rule that was used to obtain the exact centroids, and thus no further conclusion should be drawn from them. Finally, looking at the second row in Figure \ref{fig:scores_wu10}, it is seen that the $\partial^1 \zeta \ \partial^1 H$ weight approximation (the one previously proposed in the literature) consistently shows a noticeable negative bias, which however improves with increasing $N_\mathrm{eff}$. All the other weight approximations show smaller biases, falling mostly below the 5\% level, with very little dependence on $N_\mathrm{eff}$. Of these, the $\partial^3 \zeta \ \partial^\infty H$ and $\partial^2 \zeta \ \partial^1 H$ approximations show negative biases, while the $\max\left( \partial^3 \zeta \ \partial^2 H \mathrm{(nd)}, \ \partial^1 \zeta \ \partial^1 H \right)$ approximation shows positive ones, for all node distributions considered.

\subsection{The most suitable approximations}
As previously mentioned, the goal of this section is to establish the truncation order of the Taylor-series expansions that yields the most accurate weight and centroid approximations with the least amount of computational cost and complexity. With regards to accuracy, interpreting it in terms of error dispersion and bias, there is no clear winner between the $\partial^2 \zeta \ \partial^1 H$ and $\max\left( \partial^3 \zeta \ \partial^2 H \mathrm{(nd)}, \ \partial^1 \zeta \ \partial^1 H \right)$ weight approximations. The same is true for all the shortlisted centroid approximations. However, due to the lower expansion order from which they emerge, computational cost and complexity clearly favour the $\partial^2 \zeta \ \partial^1 H$ weight and centroid approximations, and thus they should be the preferred choice. In spite of this, in applications and/or numerical schemes that are sensitive to weight underestimation, the $\max\left( \partial^3 \zeta \ \partial^2 H \mathrm{(nd)}, \ \partial^1 \zeta \ \partial^1 H \right)$ weight approximation can be worthy of consideration.

Having selected a preferred weight and centroid approximation, the final figure in this section, Figure \ref{fig:scores_all}, shows a comparison of the kernels mentioned in Section \ref{sec:thekernels}. As expected, kernels that are less centrally concentrated consistently yield better results. Naturally, this holds for all the approximations, not just the $\partial^2 \zeta \ \partial^1 H$ one.

\section{Properties of the cubature rule}
\label{sec:propcb}
As previously mentioned, the partition-of-unity-based cubature rule of \gls{awph} plays the role of a spatial discretisation in a scheme to solve a conservation law. Naturally, it is but one of several ingredients that need to be put together in order to obtain a practical scheme, and each of these ingredients will make its contribution to the overall error of the scheme. This section aims at establishing empirically the basic properties of the cubature-rule's contribution to this error budget. Additionally, the tests conducted provide the opportunity to validate the results of Section \ref{sec:kdemci}. Consequently, first a likely lower bound on the rule's error is established by considering the perhaps trivial case of integrating a constant function under a regular node distribution. Then, more general node distributions are considered for the integration of some basic functions. Throughout this section, for the sake of brevity, the weight approximation $\max \left( \partial^3 \zeta \ \partial^2 H \mathrm{(nd)}, \ \partial^1 \zeta \ \partial^1 H \right)$ will be referred to simply as $\partial^3 \zeta \ \partial^2 H$. Lastly, all tests are carried out in three spatial dimensions.

\subsection{A regular node distribution}

\begin{figure}
    \centering
    \includegraphics[width=\linewidth,keepaspectratio]{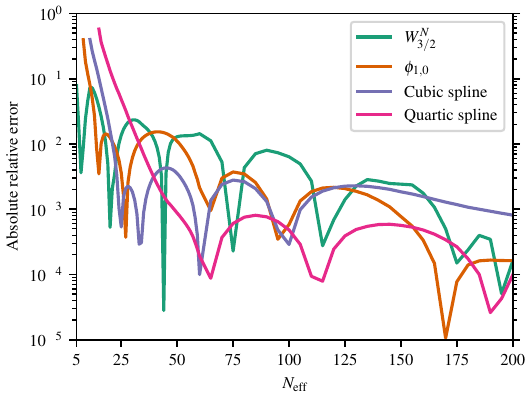}
    \caption{Relative error as a function of $N_\mathrm{eff}$ for the integration of a constant function when the nodes are distributed in a cubic lattice. A total of $N = 10^3$ nodes are distributed in a unit cube. Results are shown for four kernel functions, as indicated.}
    \label{fig:cubic}
\end{figure}

Given that the weights in the cubature rule are approximated, the latter will not be exact even for the integration of constant functions, irrespective of the node distribution. Thus, a likely lower bound on its error (or, equivalently, an upper bound on its accuracy) can be placed with the integration of a constant function under a regular node distribution. A cubic lattice is used for the latter; however, conclusions remain unchanged if a face-centred cubic lattice or a so-called glass-like distribution \citep{w96} are used instead.

It should be noted that the support function is not a constant under such node distributions. Nonetheless, it is indeed the case that, due to the node regularity, it evaluates to the same value at all node positions, and furthermore its gradient and the off-diagonal entries of its Hessian matrix all vanish at node positions. The Hessian diagonal entries, however, do not. Thus, all weight approximations that do not depend on the diagonal entries of the support-function's Hessian matrix will yield the same result under regular node distributions, which is the case for all the approximations considered here, given the results of Section \ref{sec:propwc}.

With the above in mind, regular node distributions are a trivial case in the sense that the error in the cubature rule becomes equivalent to the error in the \gls{kde} estimate at node positions, which under such distributions is given entirely by the shape of the kernel being used. In fact, given that in this case the correct answer for the node weights is simply the size of the domain divided by the number of nodes, the error in the cubature rule is equivalent to the degree to which a given kernel can yield the ``correct'' node support, which from Equation \ref{eq:awphh} is
\begin{equation}
    H = \left( \dfrac{\left| \Omega \right|}{N} \dfrac{N_\mathrm{eff}}{C_\nu} \right)^{1/\nu},
\end{equation}
\noindent where $\left| \Omega \right|$ is the size of the domain. Thus if the kernel, through Equation \ref{eq:awphh} and for a given $N_\mathrm{eff}$, yields a support that underestimates this ``correct'' value, then the cubature rule will underestimate the value of the integral, and equivalently for the case of an overestimation. Furthermore, it should be noted that, in this case, altering the number of nodes while keeping $N_\mathrm{eff}$ constant leads to identical results, as this merely re-scales the problem, leaving relative errors untouched.

The cubature-rule's relative error as a function of $N_\mathrm{eff}$ is shown in Figure \ref{fig:cubic}, obtained with a total of $10^3$ nodes distributed as a cubic lattice in a unit cube. It shows that the likely minimum relative error that can be expected from the cubature rule is in the order of $10^{-3} \! \sim \! 10^{-4}$ for the more common kernels, namely the cubic and quartic splines, at their typical $N_\mathrm{eff}$ values. However, the $W_{3/2}^N$ kernel of \citet{c08} and the $\phi_{1,0}$ kernel of \citet{wu95} show somewhat higher relative errors at low $N_\mathrm{eff}$ values (similar to those for the cubic spline, for example), falling mostly in the $10^{-2} \! \sim \! 10^{-3}$ range. The erratic behaviour observed comes from the weight (or support, as noted above) switching back and forth from under- to overestimating the exact value, which, as noted earlier, is entirely given by the shape of the kernel. Finally, note that Figure \ref{fig:cubic} is in fact equivalent to Figure 3 in \citet{da12}, where it is used to assess the accuracy of the \gls{sph} density estimate for a regular node distribution.

\subsection{More realistic node distributions}
Having studied the trivial case of a constant function under a regular node distribution, the next step is to consider basic functions under more general node distributions. The analysis is restricted to non-negative functions and the nodes are assumed to be samples stochastically drawn from these functions after being normalised. Justification for this is provided for the case in which the conservation law to solve is the system of equations describing inviscid compressible hydrodynamics.

Recalling, the cubature rule in \gls{awph} evolves with time in order to adapt to the evolution of the fields it integrates. In the context of ideal hydrodynamics, these fields are mass density, momentum density and total-energy density. Given that the nodes are prescribed to move with the local fluid velocity in both \gls{wpm} and \gls{awph}, if equal-mass nodes are assumed and mass exchanges between them are neglected, then they are expected to trace the mass-density field of the problem being solved. Moreover, whereas in \gls{sph} some semblance of node order has been argued to be maintained due to the zeroth-order inconsistency in its gradient estimator \citep[e.g.][Section 5.1.2]{da12}, no such inconsistency is present in the \gls{rmd} estimator used in \gls{awph}. Therefore, it can be considered that node positions in \gls{awph} evolve in such a way that they always resemble statistical samples drawn from a distribution given by the normalised mass-density field of the problem being solved, thus justifying the assumptions made for the analysis in this section.

Provided such assumptions hold, there is in fact a nice result that establishes the asymptotic (in the limit of ``large'' $N$) equivalence between the bias in weight approximation and the relative error in \gls{awph}['s] cubature rule, since in that case each term in the cubature-rule's summation is asymptotically the probability mass of the corresponding node. If it can be assumed that the sampling procedure draws samples that asymptotically have the same probability mass, then each of these terms should approach that probability mass. This can then be recast into showing that weight-approximation bias and cubature-rule relative error are, under these assumptions, asymptotically equivalent.

With this in mind, and in order to mirror the analysis of Section \ref{sec:propwc}, the same distributions considered there are used here; namely, the uniform, piecewise-uniform and normal distributions (see Section \ref{sec:nodedist}). However, results for the piecewise-uniform distribution lie somewhat in between those of the uniform and normal distributions, thus only the latter two will be shown. Finally, the weight approximations being considered, as per the results of Section \ref{sec:propwc}, are the $\partial^2 \zeta \ \partial^1 H$ and $\partial^3 \zeta \ \partial^2 H$ approximations, plus the one previously proposed in the literature \citep{gn11}, namely the $\partial^1 \zeta \ \partial^1 H$ approximation (Equation \ref{eq:awphweight}).

\subsubsection{The uniform distribution}

\begin{figure}
    \centering
    \includegraphics[width=\linewidth,keepaspectratio]{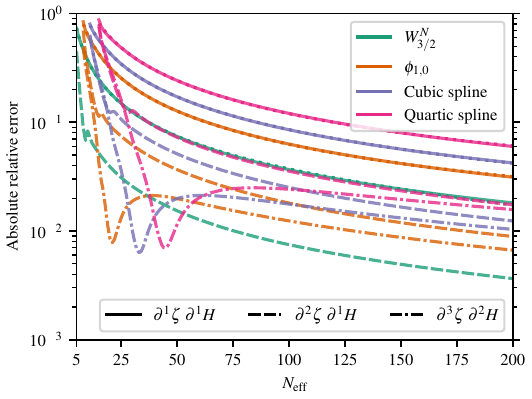}
    \caption{Relative error as a function of $N_\mathrm{eff}$ for the integration of the uniform distribution. Positions of $10^3$ nodes are drawn randomly from said distribution and their weights are computed using the kernels and approximations indicated in the legends. Dotted lines show a least-squares fit to Equation \ref{eq:meanrelerr}, for each kernel.}
    \label{fig:uniform}
\end{figure}

The relative error as a function of $N_\mathrm{eff}$ is shown in Figure \ref{fig:uniform}, in which $10^3$ node positions were drawn from the uniform distribution. In it, dotted lines show a least-squares fit of the $\Theta$, $\Phi$ and $\Xi$ coefficients of Equation \ref{eq:meanrelerr} to the relative errors corresponding to the $\partial^1 \zeta \ \partial^1 H$ weight approximation. It is seen that the best-fit coefficients have values consistent with their theoretical form (see \ref{ap:relerr}) for a uniform distribution, as $\Theta$ and $\Phi$ have vanishing values and only the $\Xi$ term remains. Thus, in this case, increasing $N$ while keeping $N_\mathrm{eff}$ constant leads to identical results. This property, however, is seen to hold not only for the $\partial^1 \zeta \ \partial^1 H$ weight approximation, but for the others as well.

Both the $\partial^2 \zeta \ \partial^1 H$ and $\partial^3 \zeta \ \partial^2 H$ approximations are seen to perform much better than the $\partial^1 \zeta \ \partial^1 H$ approximation, as their relative errors are smaller by a factor of roughly 3.5. This difference, as previously mentioned, is a direct consequence of the latter's negative bias with which it estimates weights, as seen in Figure \ref{fig:scores_wu10}, which leads directly to an underestimate of the true value of the integral. In the case of the $\partial^2 \zeta \ \partial^1 H$ approximation, the negative bias is much smaller, yet still present, thus the much smaller underestimate it yields for the integral. Finally, the $\partial^3 \zeta \ \partial^2 H$ approximation, albeit barely noticeable in Figure \ref{fig:scores_wu10}, in fact switches from a negative to a positive bias at low $N_\mathrm{eff}$ values, thus explaining why the same behaviour is observed in its relative error in Figure \ref{fig:uniform}.

\subsubsection{The normal distribution}

\begin{figure}
    \centering
    \includegraphics[width=\linewidth,keepaspectratio]{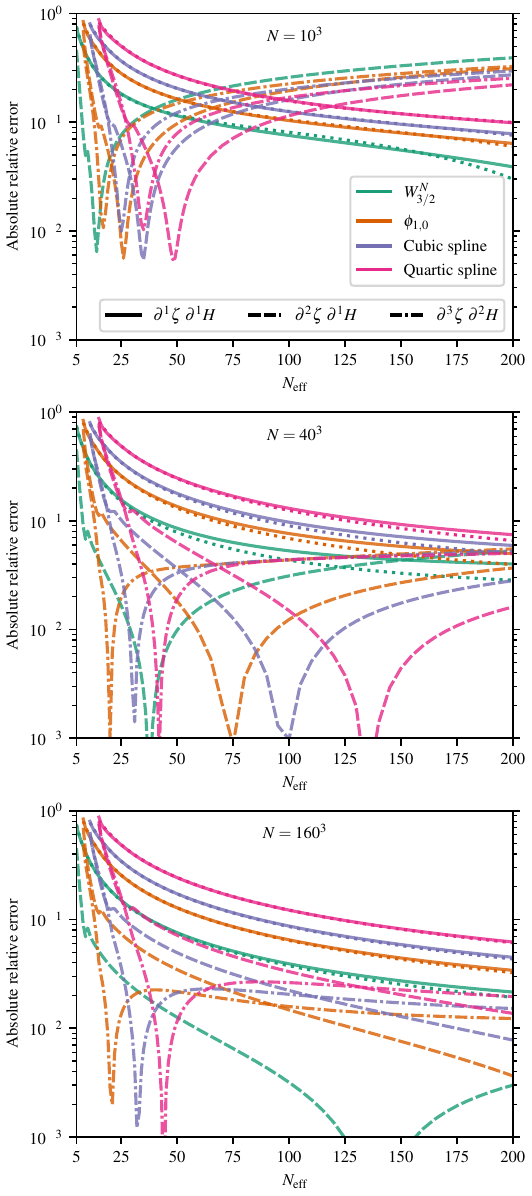}
    \caption{Relative error as a function of $N_\mathrm{eff}$ for the integration of the normal distribution with a standard deviation of 0.1. Node positions are drawn randomly from said distribution and their weights are computed using the kernels and approximations indicated in the legends. Dotted lines show a least-squares fit to Equation \ref{eq:meanrelerr}, for each kernel. From top to bottom, the number of nodes are $10^3$, $40^3$ and $160^3$.}
    \label{fig:normal}
\end{figure}

Moving on to the case of the normal distribution, Figure \ref{fig:normal} shows the relative error as a function of $N_\mathrm{eff}$ for three different numbers of nodes. As in the uniform-distribution case, dotted lines show a least-squares fit to Equation \ref{eq:meanrelerr}. As expected, in this case the best-fit values for the coefficients $\Theta$ and $\Phi$ are non-negligible; however, compared to the best-fit $\Xi$ coefficient, they are still subdominant. This means that the relative error in the $\partial^1 \zeta \ \partial^1 H$ approximation for the integration of the normal distribution is dominated by the variance of the \gls{kde}, which in turn means that the error will show little dependence on $N$, and furthermore will decrease with increasing $N_\mathrm{eff}$ up to large values of the latter before the bias and squared-bias terms begin to dominate, as Figure \ref{fig:normal} shows.

In contrast, the relative error in the $\partial^2 \zeta \ \partial^1 H$ and $\partial^3 \zeta \ \partial^2 H$ approximations show a markedly different behaviour with both $N$ and $N_\mathrm{eff}$. On one hand, for $N = 10^3$ and at large $N_\mathrm{eff}$ values, both perform very poorly; however, their performance progressively improves with increasing $N$. This poor performance is basically a consequence of nodes having overly broad supports, which renders the Taylor-series-based approach from which these approximations were derived very unfitting. On the other hand, much like the analysis in the uniform-distribution case, the behaviour of their relative errors can again be understood on the basis of the bias of their weight estimates. However, in the present case of a normal distribution, their asymptotic equivalence needs to be taken with care as finite-sample effects are more prominent. Thus, albeit approximately, the error in the integral estimate reflects the bias in the weight approximation, and the behaviour seen for the former in Figure \ref{fig:normal} basically showcases the $N_\mathrm{eff}$ values at which the respective weight approximations switch from negative to positive bias. In other words, at these $N_\mathrm{eff}$ values the weight approximations are unbiased, and they are seen to depend strongly on $N$ in the $\partial^2 \zeta \ \partial^1 H$ approximation, yet are roughly constant in the $\partial^3 \zeta \ \partial^2 H$ approximation. In fact, in the case of the latter, these $N_\mathrm{eff}$ values are similar as well in the uniform-distribution case, as seen in Figure \ref{fig:uniform}, and likewise in the piecewise-uniform distribution, although the latter is not shown.

It is thus seen that the $\partial^3 \zeta \ \partial^2 H$ weight approximation, in the simple cases here considered, can lead to a very accurate cubature rule in a narrow range of kernel-dependent values of the $N_\mathrm{eff}$ parameter. However, there is at the moment no reason to expect this nice property to be preserved for a wider class of functions.

\section{An example application}
\label{sec:sedov}

\begin{figure*}
    \centering
    \includegraphics[width=\textwidth,keepaspectratio]{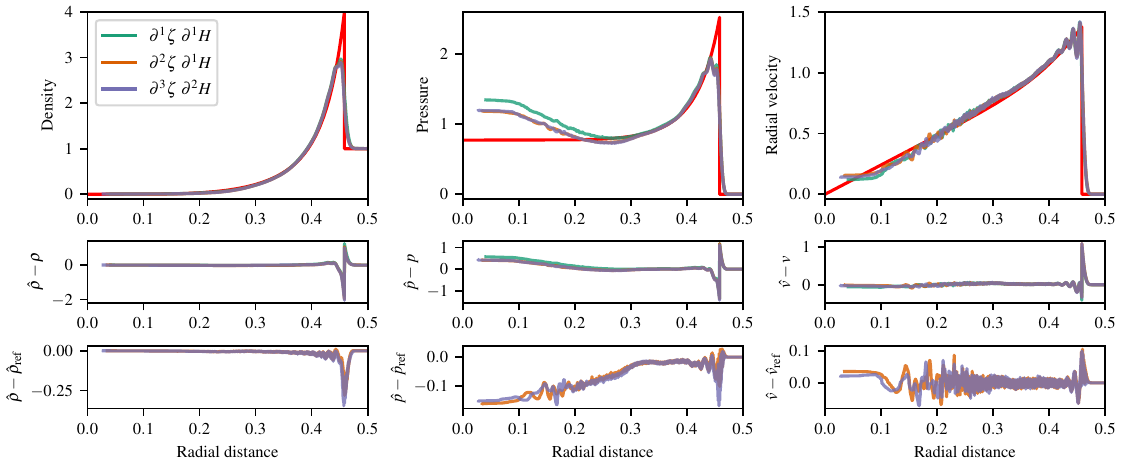}
    \caption{Properties of the Sedov-blast-wave solution at $t = 0.1$ for the three weight approximations indicated in the legend. The top row shows the radial profiles of density, pressure and radial velocity, in which the red line corresponds to the self-similar solution. The middle row shows the respective errors, and the bottom row the differences between the $\partial^2 \zeta \ \partial^1 H$ and the $\partial^3 \zeta \ \partial^2 H$ approximations with respect to the $\partial^1 \zeta \ \partial^1 H$ approximation. The $\phi_{1,0}$ kernel of \citet{wu95} was used with $N_\mathrm{eff} = 60$. All the curves were obtained by smoothing per-node values with a Gaussian filter.}
    \label{fig:sedov}
\end{figure*}

The previous two sections examined the properties of different weight approximations and their impact in \gls{awph}['s] cubature rule. As noted earlier, the cubature rule is but one of several ingredients that are required to set up a numerical scheme that fully solves a conservation law, and the scheme's overall error comes from compounding those of each individual ingredient. Furthermore, thus far node positions were considered to behave as statistical samples drawn from the function whose integral is being discretised; however, this is a pessimistic assumption in cases were nodes are expected to move in a somewhat ordered fashion. Thus, despite the differences that were observed in the cubature rule under different weight approximations, their impact on the final numerical solution will be strongly influenced by the overall numerical scheme and by the particular problem the conservation law is being used to model. With this in mind, this section aims to showcase solving an example conservation law using different weight approximations. This is additionally used to assert the preservation of the numerical-stability and conservation properties of schemes that use the weight approximation previously proposed in the literature \citep{gn11} when the latter is replaced by the ones introduced in this work. As in the previous section, for the sake of brevity, the $\max \left( \partial^3 \zeta \ \partial^2 H \mathrm{(nd)}, \ \partial^1 \zeta \ \partial^1 H \right)$ weight approximation will be referred to simply as $\partial^3 \zeta \ \partial^2 H$.

The conservation law that is considered is the system describing inviscid compressible hydrodynamics in three spatial dimensions. With it, the particular problem that is solved is the Sedov blast wave \citep{s46}, which models a point explosion at the origin of a spherically symmetric medium at rest. This problem is set up such that a supersonic shock wave develops and propagates radially through the medium, and for which a self-similar closed-form solution has been shown to exist \citep{s46,k00}. As a consequence, the problem probes a scheme's ability to correctly transform thermal energy to kinetic energy, and vice versa, in a conservative way, while conserving angular momentum and capturing a strong shock. As such, it is deemed a sensible choice for the goals of this section. The setup chosen mirrors that typically used in astrophysical applications. In arbitrary units, the medium is initially of uniform density equal to unity, vanishingly small pressure equal to $10^{-6}$, and an adiabatic index $\gamma = 5/3$. A thermal energy $U = 1$ is added at $t = 0$ to a single node residing at the origin, and the integration is carried out up to $t = 0.1$. Finally, nodes are initially distributed in a glass-like distribution with $N = 128^3$.

The numerical scheme that is used closely resembles the \gls{awph} scheme of \citet{gn11} and the subsequent contribution of \citet{h15}. It is fully described in \citet{a23}, and here only the major details are given. The scheme uses the \gls{rmd} together with the cubature rule to transform the initial weak form of the system of conservation laws into a system of ordinary differential equations for the temporal evolution of the conserved variables that each node carries. As it turns out, the final form of the latter is identical to that obtained with a finite-volume discretisation, as the temporal evolution is controlled by the exchange of fluxes through an interaction interface between neighbouring nodes. Thus, the Godunov approach is used, whereby numerical fluxes are obtained by evaluating the physical fluxes with the solution of a one-dimensional Riemann problem defined at the interface of each pair of neighbouring nodes. More specifically, the MUSCL-Hancock scheme \citep{v79,v82} is used, by which the fluid states at each side of an interface are obtained by linearly extrapolating in space and time the fluid states of the corresponding nodes. After accumulating the fluxes exchanged by each node, a simple Euler step updates their conserved variables in time. Finally, \citet{h15} introduced two versions of the scheme, namely the meshless-finite-volume and meshless-finite-mass schemes. They differ in the characteristic along which the solution to the Riemann problem is sampled at each interface. In the former, a weighted average of the fluid velocities of the corresponding nodes is used, while in the latter the solution is always sampled along the contact discontinuity. Results are shown only for the meshless-finite-mass scheme, however, as the conclusions are the same for either.

Figure \ref{fig:sedov} shows the results obtained at $t = 0.1$ using the $\phi_{1,0}$ kernel of \citet{wu95} with $N_\mathrm{eff}$ = 60. It can be seen that there are only minor differences between the results obtained with the $\partial^1 \zeta \ \partial^1 H$, $\partial^2 \zeta \ \partial^1 H$ and $\partial^3 \zeta \ \partial^2 H$ weight approximations. The most notable difference is in the pressure well behind the shock, where the results for the $\partial^2 \zeta \ \partial^1 H$ and $\partial^3 \zeta \ \partial^2 H$ approximations are somewhat closer to the self-similar solution, when compared to the results for the $\partial^1 \zeta \ \partial^1 H$ approximation. Additionally, the lower-left panel shows that the $\partial^2 \zeta \ \partial^1 H$ and $\partial^3 \zeta \ \partial^2 H$ approximations are consistently yielding lower densities when compared to the $\partial^1 \zeta \ \partial^1 H$ approximation. In fact, in this problem it is seen that the node distributions at all stages of the evolution are very similar between all the weight approximations, thus this density underestimation is yet another manifestation of the negative bias in the $\partial^1 \zeta \ \partial^1 H$ weight approximation, which leads to nodes having higher densities for a fixed node distribution when compared to the other two approximations. The differences, however, are mostly seen at the shock front, and the reason is the following. Since this problem involves a spherically symmetric flow, there is no shearing component in the motion. As a consequence, the initially regular node distribution is mostly radially expanded around the location of the explosion, without being significantly distorted. Thus, in this problem the differences between weight approximations are mostly due to node clustering, instead of node disorder. The former is, quite naturally, occurring at the shock front, thus the bulk of the differences in density concentrating there. However, this density difference is not large enough to significantly impact the evolution, as obtained by the scheme being used.

Finally, although not shown, it has been verified that the evolution of the (numerical) conversion of thermal to kinetic energy is virtually the same for all weight approximations, and that the scheme's conservation properties, which are detailed in \citet{a23}, are unchanged up to floating-point precision.

\section{Discussion}
\label{sec:discussion}
\subsection{Weight approximations}
The results of Section \ref{sec:propwc} show that the weight approximation proposed by \citet{gn11} has a notable negative bias, which is to say that, on average, it clearly underestimates the true weights. This bias is seen to be a strong function of the $N_\mathrm{eff}$ parameter, and is mostly in the 10-30\% range for the values that are commonly used with the cubic- and quartic-spline kernels. These percentages, however, correspond to cases for which node positions can be considered to behave as random samples drawn from the function whose integral is being discretised, and Section \ref{sec:propcb} provides a justification for this assumption for the case of inviscid compressible hydrodynamics. Thus, the actual bias observed in a given application can likely be very different from the 10-30\% mentioned above, for better or for worse. For instance, in cases where some node regularity is expected to be maintained, the bias will naturally be smaller. In contrast, in cases where the nodes can get significantly clustered, either due to the characteristics of the problem being solved or due to numerical artefacts, the bias will likely be larger.

Regarding the impact that the above-mentioned bias has on the accuracy of \gls{awph}['s] cubature rule, Section \ref{sec:propcb} shows that, under the assumptions made, it directly leads to an underestimate of the true value of an integral. Thus, the cubature rule inherits the 10-30\% error that is present in this weight approximation, once again provided that node positions can be considered to be randomly distributed. Furthermore, results from Section \ref{sec:kdemci} indicate that the rule's asymptotic convergence rate goes as $N^{-2/3}$, provided that the bias term in Equation \ref{eq:meanrelerr} is dominant. However, in the simple cases considered in Section \ref{sec:propcb} this is not seen to occur, as the finite-sample convergence properties are in fact very poor, indicating that in those cases the variance term is the one that is dominant.

In contrast, the weight approximation that emerged as the most suitable from the analysis in Section \ref{sec:propwc}, the $\partial^2 \zeta \ \partial^1 H$ approximation, has notably better properties as compared to the one proposed by \citet{gn11}. Its bias is much reduced, being mostly below the 5\% level, which likewise leads to the cubature rule displaying a similar accuracy in the simple cases considered in Section \ref{sec:propcb}. Furthermore, the rule's accuracy becomes a more complicated function of $N$ and $N_\mathrm{eff}$, as seen in Figure \ref{fig:normal}, showing that, with this weight approximation, increasing $N$ while keeping $N_\mathrm{eff}$ constant can in fact be beneficial.

The results obtained for the Sedov-blast-wave problem in Section \ref{sec:sedov} reinforce the previous points, in the sense that the difference between weight approximations is most evident in regions of node disorder and/or node clustering. Given that the Sedov blast wave involves a spherically symmetric flow, thus with no shearing component, node clustering is the most impactful in this case and, naturally, the latter occurs at the shock front. However, the dynamical impact of this difference is seen to be minor in this case, as the most notable effect is a small difference in the pressure profile in the region way behind the shock front. This highlights, once again, that the better properties displayed by the weight approximations derived in this work will not necessarily translate into improvements on the final solution of a conservation law. This will depend, on the one hand, on the problem being solved (more specifically, on the node disorder and clustering that are expected during its evolution) and, on the other hand, on the remaining details of the numerical scheme being used.

\subsection{Centroid approximations}
Even though centroid approximations are derived and compared to their exact calculation, putting them to test in a practical set up is beyond the scope of the present work. Notwithstanding, their practical significance and applicability is briefly discussed in the following.

In the present context, centroids are mainly indicators of the first-order errors in the numerical estimation of an integral. Indeed, as it was mentioned previously, centroidal node distributions lead to second-order-accurate cubature rules (that is, rules that are exact for affine functions). In schemes like \gls{awph}, however, centroids can only be approximated, thus their usage will be profitable only if approximations of the required accuracy are available. In the case of the approximations derived in this work, Section \ref{sec:propwc} shows that, when the nodes can be considered to be randomly distributed, they have a fair amount of scatter. As such, it is perhaps unlikely that their use will lead to more accurate results. However, in cases where nodes can be expected to maintain some degree of regularity, these approximations can prove to be useful.

It should be mentioned that the use of arbitrarily accurate cubature rules, such as those in the \textsc{Cuba} library\footnote{\url{https://feynarts.de/cuba}}, is an alternative worthy of consideration in cases of modest computational load and for which there are no centroid approximations of sufficient accuracy. Such could be the case, for example, when the issue at hand is generating optimal initial conditions (in the form of a centroidal node distribution) for a given problem.

\subsection{The $N_\mathrm{eff}$ parameter and the kernel function}
The results of Section \ref{sec:propwc} consistently show that, to varying degrees, the bias and dispersion in weight and centroid approximations, with respect to their exact values, improve for increasing $N_\mathrm{eff}$ values. The same is true, overall, regarding the accuracy of the cubature rule, as per Section \ref{sec:propcb}. However, when considering the full solution of a conservation law, there is of course an upper limit for increasing $N_\mathrm{eff}$, which is given by spatial-resolution and computational-cost considerations. Indeed, increasing $N_\mathrm{eff}$ will progressively smear out the spatial features of the solution, as the ``region of influence'' of the nodes will get progressively larger. Computational cost increases for the same reason. Thus, despite the properties displayed by the spatial discretisation, the optimal $N_\mathrm{eff}$ choice must be a compromise between all these considerations, and as such must unavoidably be determined in a case-by-case manner.

A similar conclusion can be drawn regarding the choice of kernel function. As with the $N_\mathrm{eff}$ parameter, Sections \ref{sec:propwc} and \ref{sec:propcb} clearly show that less centrally concentrated kernels perform better both in terms of weight- and centroid-approximation bias and dispersion, and in terms of accuracy of the cubature rule. The reason for this, as mentioned in Section \ref{sec:thekernels}, is that such kernels have derivatives that are smaller in magnitude, which directly lead to a support function with the same properties, thus making a Taylor-series approximation more profitable. The impact on spatial resolution and computational cost, however, is not as straightforward, since for a given $N_\mathrm{eff}$ value, more centrally concentrated kernels yield, compared to their counterparts, smaller supports in regions of high node density and larger supports in regions of low node density. Thus, it is difficult to make a general a-priori argument regarding their spatial-resolution and computational-cost properties, and as such the kernel choice should again be made in a case-by-case manner. For instance, in applications where a large number of neighbours is required, due to numerical-stability considerations for example, using a more centrally concentrated kernel provides a mechanism to increase spatial resolution while accounting for this constraint.

\subsection{The computational cost of the approximations}
Computing the analytical derivatives of the support function (Equations \ref{eq:gradh} and \ref{eq:hessh}) requires no additional loop over the nodes. They can be obtained in the same loop in which the node supports are calculated, and thus the same is true for all the weight and centroid approximations here proposed. Furthermore, throughout said loop's execution, the gradient and Hessian-matrix entries can be stored in a single temporary memory buffer. As such, the main penalty introduced by the weight and centroid approximations is an increased amount of floating-point operations required while looping over the neighbours of each node. Naturally, in the case of the $\partial^2 \zeta \ \partial^1 H$ approximations, only the gradient needs to be computed, in which case the increment is minor compared to the operations that are already being carried out, especially those required for the \gls{rmd}.

\section{Summary and conclusions}
\label{sec:conclusions}
The present work introduces new weight and centroid approximations for the partition-of-unity-based cubature rule of the \gls{awph} method. Together with the weight approximation previously proposed in the literature, they are then evaluated by comparing them to a computationally expensive exact calculation of weights and centroids. As a result, optimal weight and centroid approximations are selected on the basis of their accuracy and computational cost and complexity. These turn out to be the approximations arising from a second-order Taylor-series expansion of the kernel function and a first-order Taylor-series expansion of the support function; namely, the ones labelled as $\partial^2 \zeta \ \partial^1 H$ throughout this work.

This evaluation process revealed that the weight approximation previously proposed in the literature is, on average, a clear underestimate of the exact weights. This underestimation is seen to be a strong function of the $N_\mathrm{eff}$ parameter, such that it is reduced with increasing $N_\mathrm{eff}$; nonetheless, it lies mostly in the 10-30\% range for the values of the latter that are commonly used with the cubic- and quartic-spline kernels. In contrast, the $\partial^2 \zeta \ \partial^1 H$ approximation, while still on average underestimating the exact weights, does so at mostly below the 5\% level, and with a smaller dependence on the $N_\mathrm{eff}$ parameter.

Additionally, the accuracy and convergence of \gls{awph}['s] cubature rule is studied both theoretically and numerically. On the theoretical side, it is established that the cubature rule, with weights computed as previously proposed in the literature, is equivalent to an importance-sampling Monte Carlo integration scheme in which a kernel density estimate of the node-position distribution is used as the importance-sampling distribution. With this, an asymptotic expression for the mean relative error of the cubature rule is derived, which has contributions coming from the bias, squared bias and variance of the kernel density estimate. It indicates that the cubature rule inherits the bias-variance trade-off from the kernel density estimate, as this asymptotic mean-relative-error expression is not a monotonic function of the $N_\mathrm{eff}$ parameter. It also predicts that the rule can have very poor convergence properties in situations where the variance term dominates the overall error. Furthermore, it is shown that this asymptotic expression can be used to interpret the finite-sample properties of the cubature rule, as it is successfully used in Section \ref{sec:propcb} as a regression model for the relative errors obtained numerically in the integration of some simple functions.

Extending the analysis of the cubature rule to weight approximations different from the one proposed by \citet{gn11}, however, requires a numerical approach. Such an approach shows that the weight approximations derived in this work can have a significant impact in the accuracy of the cubature rule. Due to the assumptions made, this impact is seen to be asymptotically equivalent to the bias in the weight approximations. As a consequence, the improvements in the accuracy of the cubature rule are very similar (up to the validity of such asymptotic equivalence for a finite sample) to those in the bias of weight estimation. As such, the weight approximation previously proposed in the literature leads to a cubature rule that consistently underestimates the value of an integral, with a factor that is mostly in the range of 10-30\%. Similarly, the $\partial^2 \zeta \ \partial^1 H$ weight approximation leads to a cubature rule with an error that is mostly below the 5\% level.

It is important to mention that the validity of these results depends crucially on that of the assumptions made. The most important is that node positions in \gls{awph} can be considered to behave as random samples drawn from the function whose integral is being discretised, and Section \ref{sec:propcb} provides a justification for it for the case of inviscid compressible hydrodynamics. Furthermore, given that the spatial discretisation is but one of the ingredients needed to obtain a scheme that fully solves a conservation law, and that each of them makes its own contribution to the overall error, the impact of the approximations here proposed will additionally depend on all of these other factors. Section \ref{sec:sedov} showcases this by solving the Sedov-blast-wave problem of inviscid compressible hydrodynamics using different weight approximations. In it, for the numerical scheme that is used, it is seen that there are only minor differences in the solutions obtained with each weight approximation. This is the case because, in this problem, a spherically symmetric flow develops, which mostly radially expands the initially regular node distribution, yet not significantly distorting it. As such, the differences between the weight approximations are mostly due to node clustering, which in this problem occurs at the shock front. However, the differences there are not large enough to significantly disrupt the numerical solution as obtained by the scheme that is used, and thus the minor changes that are observed.

Evidently, then, whether or not the better properties displayed by the weight approximations derived in this work will translate into improvements on the final solution of a conservation law depends on the problem being solved (more specifically, on the presence of node clustering and/or disorder) and the numerical scheme being used. Regarding the former, in the context of fluid-flow problems, this is expected to be the case in the presence of large density gradients and/or strong shocks, in low-Mach-number flows (in which the dynamics is dominated by density variations), and in flows with a strong shearing component. Regarding the latter, it is to be expected that higher-order schemes, given their smaller truncation errors, are more likely to exhibit a stronger response.

In conclusion, while introducing new approximations for node weights and centroids for \gls{awph}, this work attempts to lay the groundwork for their consideration in the different applications that the method has. It is hoped that enough information is provided such that the users of the method can make an informed decision regarding their relevance for a given application.

\printacronyms

\section*{Acknowledgements}
To the anonymous referee, whose comments and suggestions contributed to the improvement of the manuscript.

This work is part of grant CEX2025-001609-S, awarded to the Instituto de Astrofísica de
Canarias under the Severo Ochoa Centre of Excellence program and funded by
MICIU/AEI/10.13039/501100011033. It has been further supported by the Spanish Ministry of Science, Innovation and Universities (Ministerio de Ciencia, Innovación y Universidades, MICIU) through the research grants PGC2018-094975-B-C22 with reference PRE2019-088629 and PID2021-122603NB-C22.

The authors wish to further acknowledge the contribution of Teide High-Performance Computing facilities and Instituto de Astrofísica de Canarias (IAC) High-Performance Computing support team and hardware facilities to the results of this research. TeideHPC facilities are provided by the Instituto Tecnológico y de Energías Renovables (ITER, SA). URL: https://teidehpc.iter.es.

Finally, extensive use was made of the following Python packages: NumPy \citep{numpy}, SciPy \citep{scipy} and Matplotlib \citep{mpl}.

\appendix

\section{The asymptotic relative error of \gls{awph}['s] cubature rule}
\label{ap:relerr}
As seen in Section \ref{sec:kdemci}, \gls{awph}['s] cubature rule, with weights computed using the approximation proposed by \citet{gn11}, is equivalent to an importance-sampling \gls{mci} scheme with a \gls{kde} as the importance-sampling distribution. An asymptotic expression (in the sense described in the following) for the mean relative error of such a scheme can be obtained by resorting to results from \gls{kde} theory.

The scheme aims to approximate the integral $I = \int F(\myvec{x}) \, d\myvec{x}$ with the random variable
\begin{equation}
    \hat{I} = \dfrac{1}{N} \sum_{i=1}^{N}\dfrac{F(\myvec{X}_i)}{\hat{f}(\myvec{X}_i)}.
    \label{eq:ihat}
\end{equation}
\noindent It is assumed that the function $F(\myvec{x})$ is non-negative and that node positions $\myvec{X}_i$ are samples of $f(\myvec{x}) = F(\myvec{x}) / I$. The \gls{kde}, which is a random variable as well, is given by
\begin{equation}
    \hat{f}(\myvec{x}) = \dfrac{1}{N} \sum_{j=1}^N W(|\myvec{x}-\myvec{Y}_j|,H(\myvec{x})),
\end{equation}
\noindent where $W(x,H)$ is the kernel function and the random vectors $\myvec{Y}_i$ are again the node positions. The use of both $\myvec{X}_i$ and $\myvec{Y}_i$ to denote node positions is needed to separate the two stochastic measurements being made, namely those of $\hat{I}$ and $\hat{f}(\myvec{x})$. In general, $\myvec{X}_i$ and $\myvec{Y}_i$ follow different distributions; however, in the case at hand they are the same sample drawn from $f(\myvec{x})$.

With a Taylor-series expansion, Equation \ref{eq:ihat} can be written as
\begin{equation}
    \dfrac{\hat{I}-I}{I} \approx \dfrac{1}{N} \sum_{i=1}^N \left[ \dfrac{f_i-\hat{f}_i}{f_i} + \left( \dfrac{f_i-\hat{f}_i}{f_i} \right)^2 \right],
\end{equation}
\noindent where $f_i = f(\myvec{X}_i)$ and likewise for $\hat{f}_i$. Taking the expectation with respect to $\myvec{Y}$ gives
\begin{equation}
    \mathbb{E}_\myvec{Y} \left( \dfrac{\hat{I}-I}{I} \right) \approx \dfrac{1}{N} \sum_{i=1}^N \left[ \dfrac{\mathbb{E}_\myvec{Y} \left( f_i-\hat{f}_i \right)}{f_i} + \dfrac{\mathbb{E}_\myvec{Y} \left( f_i-\hat{f}_i \right)^2}{f^2_i} \right].
    \label{eq:samplerelerr}
\end{equation}
\noindent The numerators in the terms under the summation are the (negative) bias and the mean squared error of the KDE, respectively. \gls{kde} theory provides a result for the leading terms of an asymptotic expansion of such quantities \cite[e.g.][Equations 4.9 and 4.10]{s86}
\begin{equation}
    \begin{gathered}
        \mathbb{E}_\myvec{Y}(\hat{f}-f)(\myvec{x}) \approx \dfrac{\alpha}{2} H^2(\myvec{x}) \nabla^2 f(\myvec{x}),\\
        \mathbb{E}_\myvec{Y}(\hat{f}-f)^2(\myvec{x}) \approx \dfrac{\alpha^2}{4} H^4(\myvec{x}) \left( \nabla^2 f(\myvec{x}) \right)^2 + \beta \dfrac{f(\myvec{x})}{N H^\nu(\myvec{x})}, \label{eq:asympt}
    \end{gathered}
\end{equation}
\noindent where $\alpha$ and $\beta$ are constants that depend on the kernel function. It should be noted, however, that these asymptotic expressions are valid for a support function, $H(\myvec{x})$, that is not a function of the data; the latter in the case at hand being the $\myvec{Y}_i$. In \gls{awph}, of course, this is not fulfilled, as Equation \ref{eq:awphh} is equivalent to
\begin{equation}
    H(\myvec{x}) = \left( \dfrac{N_\mathrm{eff}}{C_\nu N \hat{f}(\myvec{x})} \right)^{1/\nu}.
\end{equation}
\noindent One way to be able to proceed, due to \citet{ts92}, is to use the \gls{kde} result that if $N \to \infty$, $H \to 0$ and simultaneously $NH^\nu \to 0$, then $\hat{f}(\myvec{x}) \to f(\myvec{x})$. Thus,
\begin{equation}
    H(\myvec{x}) \to \left( \dfrac{N_\mathrm{eff}}{C_\nu N f(\myvec{x})} \right)^{1/\nu},
\end{equation}
\noindent Using this asymptotic value instead makes the results of Equations \ref{eq:asympt} usable. Then, substituting these in Equation \ref{eq:samplerelerr} and taking the expectation with respect to $\myvec{X}$ leads to
\begin{equation}
    \mathbb{E}\left( \dfrac{\hat{I}-I}{I} \right) \approx \Theta \left( \dfrac{N_\mathrm{eff}}{N} \right)^{2/\nu} + \Phi \left( \dfrac{N_\mathrm{eff}}{N} \right)^{4/\nu} + \dfrac{\Xi}{N_\mathrm{eff}},
    \label{eq:apmeanrelerr}
\end{equation}
\noindent in which
\begin{equation}
    \begin{gathered}
        \Theta = -\dfrac{\alpha}{2 C_\nu^{2/\nu}} \mathbb{E}_\myvec{X} \left(\dfrac{\nabla^2 f(\myvec{X})}{f^{2/\nu+1}(\myvec{X})} \right), \\
        \Phi = \dfrac{\alpha^2}{4 C_\nu^{4/\nu}} \mathbb{E}_\myvec{X} \left(\dfrac{\nabla^2f(\myvec{X})}{f^{2/\nu+1}(\myvec{X})} \right)^2, \\
        \Xi = C_\nu \beta.
    \end{gathered}
\end{equation}

Equation \ref{eq:apmeanrelerr} is the sought-out result. It gives the asymptotic scaling  with $N_\mathrm{eff}$ and $N$ of the scheme's mean relative error. Due to the assumptions made, however, care should be taken when using it to understand the finite-sample properties of \gls{awph}['s] cubature rule (once again, when its weights are computed as proposed by \citet{gn11}). For instance, in Section \ref{sec:propcb} it is successfully used as a regression model to fit the observed relative errors while numerically integrating some simple functions.

\section{The coefficients in the approximations for weights and centroids}
\label{ap:coeff}
In this appendix, the explicit form taken by the coefficients in the weight and centroid approximations derived in Section \ref{sec:derivation} are given. Recalling that the centroid is given by $\myvec{c}_i = \myvec{s}_i / w_i + \myvec{x}_i$, the approximations are written as
\begin{gather*}
    w_i \approx \dfrac{1}{\omega(\myvec{x}_i,H_i)}\left( 1 + \alpha_\zeta \Alpha_i + \beta_\zeta \Beta_i + \gamma_\zeta \Gamma_i\right), \\
    \myvec{s}_i \approx \dfrac{1}{\omega(\myvec{x}_i,H_i)} \left( \alpha_\zeta \mymat{A}_i + \beta_\zeta \mymat{B}_i + \gamma_\zeta \mymat{\Gamma}_i \right) \myvec{g}_i.
\end{gather*}
\noindent where lower- and upper-case bold-face letters denote column vectors and matrices, respectively. For the sake of conciseness, the expressions that follow are compacted using matrix operations, even though in the one-dimensional case all quantities reduce to scalars. The support function, its gradient vector and its Hessian matrix, evaluated at $\myvec{x} = \myvec{x}_i$, are denoted as $H_i$, $\myvec{g}_i$ and $\mymat{H}_i$, respectively, while $\mymat{I}$ is the identity matrix, and $\tr(\cdot)$ and $\det(\cdot)$ denote the trace and determinant operations. Evaluations of the kernel-dependent constants $\alpha_\zeta$, $\beta_\zeta$ and $\gamma_\zeta$ are given in Tables \ref{tab:kconst1d}, \ref{tab:kconst2d} and \ref{tab:kconst3d} for the one-, two- and three-dimensional cases, respectively. Kernel definitions can be found in \citet[][Equation 3]{c08}, \citet[][Table 1]{wu95} and \citet[][Table 1]{da12}.

In order to identify the order of the \gls{dk} Taylor-series expansion a given term in the expressions that follow comes from, the latter's combined power in the gradient and Hessian-matrix components must be looked at. This power is linear, quadratic or cubic if the term comes from the first-, second- or third-order Taylor-series-expansion terms, respectively. It should be noted that a gradient factor has been factored out in the centroid (that is, the $\myvec{s}_i$ vector) approximation above, thus the real power of the terms in the $\mymat{A}_i$, $\mymat{B}_i$ and $\mymat{\Gamma}_i$ coefficients is one plus that determined by looking at those terms alone. For example, the term $\tr(\mymat{H}_i)$ is linear for the weight approximation, but quadratic for the centroid. Furthermore, the weight term $\myvec{g}_i^\top \mymat{H}_i \myvec{g}_i$ is cubic, and so is the centroid term $\tr(\mymat{H}_i) \mymat{H}_i$.

\subsection{One spatial dimension}
\subparagraph{Weight approximation}
\begin{align*}
    \Alpha_i &= H_i \tr(\mymat{H}_i) + 2 \myvec{g}_i^\top \myvec{g}_i, \\
    \Beta_i &= H_i^2 \tr(\mymat{H}_i)^2 + 6 H_i \, \myvec{g}_i^\top \mymat{H}_i \myvec{g}_i, \\
    \Gamma_i &= H_i^3 \tr(\mymat{H}_i)^3.
\end{align*}

\subparagraph{Centroid approximation}
\begin{align*}
    \mymat{A}_i &= 2 H_i \idmat, \\
    \mymat{B}_i &= 4 H_i^2 \tr(\mymat{H}_i) \idmat + 4 H_i \, \myvec{g}_i^\top \myvec{g}_i \ \mymat{I}, \\
    \mymat{\Gamma}_i &= 6 H_i^3 \tr(\mymat{H}_i)^2 \idmat.
\end{align*}

\subparagraph{Kernel-dependent constants}
\begin{gather*}
    \alpha_\zeta = 3 \int_0^1 \chi^2 \, \zeta(\chi) \, d\chi, \\
    \beta_\zeta = 5 \int_0^1 \chi^4 \, \zeta(\chi) \, d\chi, \\
    \gamma_\zeta = \dfrac{35}{4} \int_0^1 \chi^6 \, \zeta(\chi) \, d\chi.
\end{gather*}

\begin{table}
    \centering
    \begin{tabular}{l c c c} 
        \hline
        Kernel          &   $\alpha_\zeta$  &  $\beta_\zeta$  &  $\gamma_\zeta$   \\ 
        \hline
        $W_{3/2}^N$     &    $0.20420308$   &  $0.11211008$   &         -         \\
        $\phi_{1,0}$    &       $3/20$      &     $9/140$     &       $1/24$      \\
        Cubic spline    &       $1/8$       &     $3/64$      &     $85/3072$     \\
        Quartic spline  &       $1/10$      &    $23/750$     &    $227/15000$    \\
        Wendland C2     &        $1/7$      &     $5/84$      &      $5/132$      \\
        Wendland C4     &       $1/10$      &     $1/33$      &     $25/1716$     \\
        Wendland C6     &       $1/13$      &     $5/273$     &     $25/3536$     \\
        \hline
    \end{tabular}
    \caption{Values of the constants $\alpha_\zeta$, $\beta_\zeta$ and $\gamma_\zeta$ for the one-dimensional case. The $W_{3/2}^N$ kernel is defined in Equation \ref{eq:sincdk} \citep[cf.][Equation 3]{c08} and the $\phi_{1,0}$ kernel in \citet[][Table 1]{wu95}. The rest are defined in \citet[][Table 1]{da12}.}
    \label{tab:kconst1d}
\end{table}

\subsection{Two spatial dimensions}
\subparagraph{Weight approximation}
\begin{align*}
    \Alpha_i &= H_i \tr(\mymat{H}_i) + 3 \myvec{g}_i^\top \myvec{g}_i, \\
    \Beta_i &= H_i^2 \left[ \tr(\mymat{H}_i)^2 + 2 \tr(\mymat{H}_i^2) \right] + 8 H_i \left[ \myvec{g}_i^\top \myvec{g}_i \tr(\mymat{H}_i) + 2 \myvec{g}_i^\top \mymat{H}_i \myvec{g}_i \right], \\
    \Gamma_i &= H_i^3 \left[ \tr(\mymat{H}_i)^3 + 4 \tr(\mymat{H}_i^3) \right].
\end{align*}

\subparagraph{Centroid approximation}
\begin{align*}
    \mymat{A}_i &= 2 H_i \idmat, \\
    \mymat{B}_i &= 4 H_i^2 \left[ \tr(\mymat{H}_i) \idmat + 2\mymat{H}_i \right] + 16 H_i \, \myvec{g}_i^\top \myvec{g}_i \ \mymat{I}, \\
    \mymat{\Gamma}_i &= 6 H_i^3 \left[ \tr(\mymat{H}_i)^2 \idmat + 4 \mymat{H}_i^2 \right].
\end{align*}

\subparagraph{Kernel-dependent constants}
\begin{gather*}
    \alpha_\zeta = 2\pi \int_0^1 \chi^3 \, \zeta(\chi) \, d\chi, \\
    \beta_\zeta = \dfrac{15\pi}{16} \int_0^1 \chi^5 \, \zeta(\chi) \, d\chi, \\
    \gamma_\zeta = \dfrac{7\pi}{8} \int_0^1 \chi^7 \, \zeta(\chi) \, d\chi.
\end{gather*}

\begin{table}
    \centering
    \begin{tabular}{l c c c} 
        \hline
        Kernel          &   $\alpha_\zeta$  &  $\beta_\zeta$  &  $\gamma_\zeta$   \\ 
        \hline
        $W_{3/2}^N$     &    $0.24343215$   &  $0.045573520$  &         -         \\
        $\phi_{1,0}$    &       $5/27$      &    $175/6336$   &     $49/4576$     \\
        Cubic spline    &      $31/196$     &  $3175/150528$  &    $511/67584$    \\
        Quartic spline  & $\dfrac{9759}{7.63 \cdot 10^4}$ & $\dfrac{947039}{6.7144 \cdot 10^7}$ & $\dfrac{169038401}{3.9567 \cdot 10^{10}}$ \\
        Wendland C2     &       $5/36$      &    $35/2112$    &     $49/9152$     \\
        Wendland C4     &       $8/66$      &    $45/4576$    &     $21/8320$     \\
        Wendland C6     &       $3/35$      &   $99/15232$    &    $143/103360$   \\
        \hline
    \end{tabular}
    \caption{Equivalent of Table \ref{tab:kconst1d} for the two-dimensional case.}
    \label{tab:kconst2d}
\end{table}

\subsection{Three spatial dimensions}
\subparagraph{Weight approximation}
\begin{align*}
    \Alpha_i &= H_i \tr(\mymat{H}_i) + 4 \myvec{g}_i^\top \myvec{g}_i, \\
    \Beta_i &= H_i^2 \left[ \tr(\mymat{H}_i)^2 + 2 \tr(\mymat{H}_i^2) \right] + 10 H_i \left[ \myvec{g}_i^\top \myvec{g}_i \tr(\mymat{H}_i) + 2 \myvec{g}_i^\top \mymat{H}_i \myvec{g}_i \right], \\
    \Gamma_i &= H_i^3 \left[ 5 \tr(\mymat{H}_i)^3 + 20 \tr(\mymat{H}_i^3) - 12 \det(\mymat{H}_i) - \Delta_i \right],
\end{align*}
\noindent where
\begin{equation*}
    \begin{split}
        \Delta_i &= 10 \left[ (\mymat{H}_i^{11})^3 + (\mymat{H}_i^{22})^3 + (\mymat{H}_i^{33})^3 \right] \\ &+ 12 \left( \mymat{H}_i^{11} \mymat{H}_i^{22} \mymat{H}_i^{33} + 4 \mymat{H}_i^{12} \mymat{H}_i^{13} \mymat{H}_i^{23} \right),
    \end{split}
\end{equation*}
\noindent in which $\mymat{H}_i^{jk}$ denotes the $jk$-th entry of the support's Hessian matrix.

\subparagraph{Centroid approximation}
\begin{align*}
    \mymat{A}_i &= 2 H_i \idmat, \\
    \mymat{B}_i &= 4 H_i^2 \left[ \tr(\mymat{H}_i) \idmat + 2\mymat{H}_i \right] + 20 H_i \, \myvec{g}_i^\top \myvec{g}_i \ \mymat{I}, \\
    \mymat{\Gamma}_i &= 6 H_i^3 \left\{ \left[ \tr(\mymat{H}_i)^2 + 2 \tr(\mymat{H}_i^2) \right] \idmat + 4\left[ \tr(\mymat{H}_i) \idmat + 2 \mymat{H}_i \right] \mymat{H}_i \right\}.
\end{align*}

\subparagraph{Kernel-dependent constants}
\begin{gather*}
    \alpha_\zeta = \dfrac{10\pi}{3} \int_0^1 \chi^4 \, \zeta(\chi) \, d\chi, \\
    \beta_\zeta = \dfrac{7\pi}{5} \int_0^1 \chi^6 \, \zeta(\chi) \, d\chi, \\
    \gamma_\zeta = \dfrac{2\pi}{5} \int_0^1 \chi^8 \, \zeta(\chi) \, d\chi.
\end{gather*}

\begin{table}
    \centering
    \begin{tabular}{l c c c} 
        \hline
        Kernel          &   $\alpha_\zeta$  &  $\beta_\zeta$  &  $\gamma_\zeta$   \\ 
        \hline
        $W_{3/2}^N$     &    $0.27450634$   &  $0.052708747$  &         -         \\
        $\phi_{1,0}$    &       $3/14$      &      $1/30$     &      $1/231$      \\
        Cubic spline    &       $3/16$      &    $17/640$     &     $31/9600$     \\
        Quartic spline  &      $23/150$     &   $227/12500$   &   $4417/2343750$  \\
        Wendland C2     &        $1/6$      &     $7/330$     &      $1/429$      \\
        Wendland C4     &       $5/39$      &     $1/78$      &      $1/884$      \\
        Wendland C6     &       $5/48$      &     $7/816$     &    $49/77520$     \\
        \hline
    \end{tabular}
    \caption{Equivalent of Table \ref{tab:kconst1d} for the three-dimensional case.}
    \label{tab:kconst3d}
\end{table}

\section{The case of a constant compact-support size}
\label{ap:consth}
The approach followed to derive weight and centroid approximations in Section \ref{sec:derivation} assumes the kernel's compact-support size is spatially adaptive through the implicit constraint given by Equation \ref{eq:awphh}. Naturally, for problems in which the compact-support size can be set to be a constant, a different approach has to be followed. This appendix details the approximations that can be obtained from one such approach.

The partition-of-unity functions given in Equation \ref{eq:awphpu} can be rewritten using the \gls{dk} (Equation \ref{eq:dk}) as
\begin{equation}
    \psi_i(\myvec{x},H) = \dfrac{\zeta(|\myvec{x}-\myvec{x}_i|,H)}{\sum_j \zeta(|\myvec{x}-\myvec{x}_j|,H)},
\end{equation}
\noindent Since $H$ is now a constant, a possibility is to expand the latter's denominator in a Taylor series around the point $\myvec{x}_i$, discard third- and higher-order terms, and proceed to evaluate Equation \ref{eq:puweight}. This leads to the following
\begin{align}
    w_i &\approx \dfrac{1}{\omega_i} \left( 1 + \alpha_\zeta \Alpha_i \right), \\
    \myvec{s}_i &\approx \dfrac{\alpha_\zeta \, \myvec{u}_i}{\omega_i},
\end{align}
\noindent where $\omega_i = \omega(\myvec{x}_i,H)$ and
\begin{align}
    A_i &= \dfrac{\ndim\!+\!2}{2}\dfrac{|\myvec{u}_i|^2}{H^2} + \mathtt{v}_i, \\
    \myvec{u}(\myvec{x}) &= -\dfrac{2}{\ndim\!+\!2}\dfrac{1}{H^{\ndim-1} \omega(\myvec{x},H)} \sum_j \dfrac{\myvec{x} - \myvec{x}_j}{|\myvec{x} - \myvec{x}_j|} \zeta_{,q}(q_j), \\
    \mathtt{v}(\myvec{x}) &= -\dfrac{1}{\ndim\!+\!2}\dfrac{1}{H^\ndim\omega(\myvec{x},H)} \sum_j\left[ \zeta_{,qq}(q_j) + (\ndim\!-\!1)\dfrac{\zeta_{,q}(q_j)}{q_j} \right],
\end{align}
\noindent in which $\ndim$ is the number of dimensions, $\zeta_{,q}(q_i) = d\zeta(q_i)/dq$, and likewise for $\zeta_{,qq}(q_i)$ (see Equation \ref{eq:dkqi} for the definition of $q$). As with the case of the support's spatial derivatives in Section \ref{sec:hderivatives}, care should be taken when evaluating $\myvec{u}_i = \myvec{u}(\myvec{x}_i)$ and $\mathtt{v}_i = \mathtt{v}(\myvec{x}_i)$, since the summations in their right-hand sides contain a term for $j=i$. Such terms must be evaluated by taking the limit when $\myvec{x} \to \myvec{x}_i$, which leads to
\begin{align}
    &\myvec{u}(\myvec{x}_i) = -\dfrac{2}{\ndim\!+\!2}\dfrac{1}{H^{\ndim-1} \omega_i} \sum_{j \ne i} \dfrac{\myvec{x}_i - \myvec{x}_j}{|\myvec{x}_i - \myvec{x}_j|} \zeta_{,q}(q_{ij}), \\
    &\mathtt{v}(\myvec{x}_i)\
    \begin{aligned}[t]
        = -\dfrac{1}{\ndim\!+\!2}\dfrac{1}{H^\ndim\omega_i} &\left\{ \sum_{j \ne i}\left[ \zeta_{,qq}(q_{ij}) + (\ndim\!-\!1)\dfrac{\zeta_{,q}(q_{ij})}{q_{ij}} \right] \right. \\ 
        &+ \left. \ndim\,\zeta_{,qq}(0) \vphantom{\sum_{j \ne i}} \right\},
    \end{aligned}
\end{align}
\noindent where $q_{ij} = |\myvec{x}_i - \myvec{x}_j|/H$. Finally, the constant $\alpha_\zeta$ is the same as that appearing in the variable-support case, and values for it are given in \ref{ap:coeff}.

\bibliographystyle{elsarticle-harv} 
\bibliography{references}

@article{lv08a,
    author={Lanson, Nathalie and Vila, Jean-Paul},
    title="{Renormalized Meshfree Schemes I: Consistency, Stability, and Hybrid Methods for Conservation Laws}",
    journal={\sjna{}},
    volume={46},
    number={4},
    pages={1912-1934},
    year={2008},
    doi={https://doi.org/10.1137/S0036142903427718}
}

@article{lv08b,
    author={Lanson, Nathalie and Vila, Jean-Paul},
    title="{Renormalized Meshfree Schemes II: Convergence for Scalar Conservation Laws}",
    journal={\sjna{}},
    volume={46},
    number={4},
    pages={1935-1964},
    year={2008},
    doi={https://doi.org/10.1137/S003614290444739X}
}

@ARTICLE{gn11,
       author = {{Gaburov}, Evghenii and {Nitadori}, Keigo},
        title = "{Astrophysical weighted particle magnetohydrodynamics}",
      journal = {\mnras{}},
         year = 2011,
        month = jun,
       volume = {414},
       number = {1},
        pages = {129-154},
          doi = {https://doi.org/10.1111/j.1365-2966.2011.18313.x}
}

@ARTICLE{h15,
       author = {{Hopkins}, Philip F.},
        title = "{A new class of accurate, mesh-free hydrodynamic simulation methods}",
      journal = {\mnras{}},
         year = 2015,
        month = jun,
       volume = {450},
       number = {1},
        pages = {53-110},
          doi = {https://doi.org/10.1093/mnras/stv195}
}

@ARTICLE{da12,
       author = {{Dehnen}, Walter and {Aly}, Hossam},
        title = "{Improving convergence in smoothed particle hydrodynamics simulations without pairing instability}",
      journal = {\mnras{}},
         year = 2012,
        month = sep,
       volume = {425},
       number = {2},
        pages = {1068-1082},
          doi = {https://doi.org/10.1111/j.1365-2966.2012.21439.x}
}

@ARTICLE{a23,
       author = {{Alonso Asensio}, Isaac and {Dalla Vecchia}, Claudio and {Potter}, Douglas and {Stadel}, Joachim},
        title = "{Mesh-free hydrodynamics in PKDGRAV3 for galaxy formation simulations}",
      journal = {\mnras{}},
         year = 2023,
        month = feb,
       volume = {519},
       number = {1},
        pages = {300-317},
          doi = {https://doi.org/10.1093/mnras/stac3447}
}

@ARTICLE{g23,
       author = {{Groth}, Frederick and {Steinwandel}, Ulrich P. and {Valentini}, Milena and {Dolag}, Klaus},
        title = "{The cosmological simulation code OPENGADGET3 - implementation of meshless finite mass}",
      journal = {\mnras{}},
         year = 2023,
        month = nov,
       volume = {526},
       number = {1},
        pages = {616-644},
          doi = {https://doi.org/10.1093/mnras/stad2717}
}

@ARTICLE{k24,
       author = {{Kajzer}, Adam and {Pozorski}, Jacek},
        title = "{On the partition noise in chosen particle weighted methods and its consequences for weakly-compressible flow models}",
      journal = {\jcp{}},
         year = 2024,
        month = feb,
       volume = {498},
        pages = {112653},
          doi = {https://doi.org/10.1016/j.jcp.2023.112653}
}

@article{v99,
    author = {{Vila}, Jean P.},
    title = {ON PARTICLE WEIGHTED METHODS AND SMOOTH PARTICLE HYDRODYNAMICS},
    journal = {\mmmas{}},
    volume = {09},
    number = {02},
    pages = {161-209},
    year = {1999},
    doi = {https://doi.org/10.1142/S0218202599000117}
}

@InProceedings{m99,
    author={{Moussa}, Ben B. and {Lanson}, N. and {Vila}, J. P.},
    title={Convergence of Meshless Methods for Conservation Laws Applications to Euler equations},
    booktitle="Hyperbolic Problems: Theory, Numerics, Applications",
    year={1999},
    publisher={Birkh{\"a}user Basel},
    pages={31-40},
    isbn={978-3-0348-8720-5}
}

@article{l01,
    author = {{Lanson}, N. and {Vila}, J. P.},
    title = {Meshless methods for conservation laws},
    journal = {\mcs{}},
    volume = {55},
    number = {4},
    pages = {493-501},
    year = {2001},
    doi = {https://doi.org/10.1016/S0378-4754(00)00285-8}
}

@ARTICLE{b96,
       author = {{Belytschko}, T.},
        title = "{Meshless methods: An overview and recent developments}",
      journal = {\cmame{}},
         year = 1996,
        month = dec,
       volume = {139},
       number = {1-4},
        pages = {3-47},
          doi = {https://doi.org/10.1016/S0045-7825(96)01078-X}
}

@ARTICLE{c02,
       author = {{Carpinteri}, A. and {Ferro}, G. and {Ventura}, G.},
        title = "{The partition of unity quadrature in meshless methods}",
      journal = {\ijnme{}},
         year = 2002,
        month = jul,
       volume = {54},
       number = {7},
        pages = {987-1006},
          doi = {https://doi.org/10.1002/nme.455}
}

@article{d02,
    author = {{Duflot}, Marc and {Nguyen-Dang}, Hung},
    title = {A truly meshless Galerkin method based on a moving least squares quadrature},
    journal = {\cnme{}},
    year = {2002},
    month = jul,
    volume = {18},
    number = {6},
    pages = {441-449},
    doi = {https://doi.org/10.1002/cnm.503}
}

@ARTICLE{g06,
       author = {{Guessab}, Allal and {Nouisser}, Otheman and {Schmeisser}, Gerhard},
        title = "{A Definiteness Theory for Cubature Formulae of Order Two}",
      journal = {\ca{}},
         year = 2006,
       volume = {24},
        pages = {263–288},
          doi = {https://doi.org/10.1007/s00365-005-0619-4}
}

@inproceedings{s68,
    author = {{Shepard}, Donald},
    title = {A two-dimensional interpolation function for irregularly-spaced data},
    year = {1968},
    isbn = {9781450374866},
    publisher = {Association for Computing Machinery},
    doi = {https://doi.org/10.1145/800186.810616},
    booktitle = {Proceedings of the 1968 23rd ACM National Conference},
    pages = {517–524},
    numpages = {8}
}

@INPROCEEDINGS{w96,
       author = {{White}, S.~D.~M.},
        title = "{Formation and Evolution of Galaxies}",
    booktitle = {Cosmology and Large Scale Structure},
    publisher = {Elsevier Science},
         year = 1996,
        month = jan,
        pages = {349},
         isbn = {0444823352}
}

@ARTICLE{s10b,
       author = {{Springel}, Volker},
        title = "{Smoothed Particle Hydrodynamics in Astrophysics}",
      journal = {\araa{}},
         year = 2010,
        month = sep,
       volume = {48},
        pages = {391-430},
          doi = {https://doi.org/10.1146/annurev-astro-081309-130914}
}

@book{e80,
    author = {{Engels}, Hermann},
    title = {Numerical quadrature and cubature},
    publisher = {Academic Press},
    year = {1980},
    isbn = {012238850X}
}

@book{d84,
    title = {Methods of Numerical Integration},
    author = {{Davis}, Philip J. and {Rabinowitz}, Philip},
    publisher = {Academic Press},
    edition = {2nd},
    year = {1984},
    doi = {https://doi.org/10.1016/C2013-0-10566-1}
}

@book{e00,
    author = {{Evans}, Michael and {Swartz}, Tim},
    title = {Approximating Integrals via Monte Carlo and Deterministic Methods},
    publisher = {Oxford University Press},
    year = {2000},
    month = {03},
    doi = {https://doi.org/10.1093/oso/9780198502784.001.0001}
}

@book{s86,
    author = {{Silverman}, Bernard Walter},
    title = {Density Estimation for Statistics and Data Analysis},
    publisher = {Chapman and Hall},
    year = {1986},
    doi = {https://doi.org/10.1201/9781315140919}
}

@article{ts92,
    author = {{Terrell}, George R. and {Scott}, David W.},
    title = {Variable Kernel Density Estimation},
    journal = {\as{}},
    volume = {20},
    number = {3},
    pages = {1236-1265},
    doi = {https://doi.org/10.1214/aos/1176348768},
    year = {1992},
    month = {September}
}

@article{wu95,
    author = {{Wu}, Zongmin},
    title = {Compactly supported positive definite radial functions},
    journal = {\acm{}},
    volume = {4},
    pages = {283–292},
    doi = {https://doi.org/10.1007/BF03177517},
    year = {1995}
}

@Article{numpy,
 title         = {Array programming with {NumPy}},
 author        = {Charles R. Harris and K. Jarrod Millman and St{\'{e}}fan J.
                 van der Walt and Ralf Gommers and Pauli Virtanen and David
                 Cournapeau and Eric Wieser and Julian Taylor and Sebastian
                 Berg and Nathaniel J. Smith and Robert Kern and Matti Picus
                 and Stephan Hoyer and Marten H. van Kerkwijk and Matthew
                 Brett and Allan Haldane and Jaime Fern{\'{a}}ndez del
                 R{\'{i}}o and Mark Wiebe and Pearu Peterson and Pierre
                 G{\'{e}}rard-Marchant and Kevin Sheppard and Tyler Reddy and
                 Warren Weckesser and Hameer Abbasi and Christoph Gohlke and
                 Travis E. Oliphant},
 year          = {2020},
 month         = sep,
 journal       = {\nature{}},
 volume        = {585},
 number        = {7825},
 pages         = {357--362},
 doi           = {https://doi.org/10.1038/s41586-020-2649-2}
}

@Article{mpl,
  Author    = {{Hunter}, John D.},
  Title     = {Matplotlib: A 2D graphics environment},
  Journal   = {\cse{}},
  Volume    = {9},
  Number    = {3},
  Pages     = {90--95},
  doi       = {https://doi.org/10.1109/MCSE.2007.55},
  year      = 2007
}

@ARTICLE{scipy,
  author  = {Virtanen, Pauli and Gommers, Ralf and Oliphant, Travis E. and
            Haberland, Matt and Reddy, Tyler and Cournapeau, David and
            Burovski, Evgeni and Peterson, Pearu and Weckesser, Warren and
            Bright, Jonathan and {van der Walt}, St{\'e}fan J. and
            Brett, Matthew and Wilson, Joshua and Millman, K. Jarrod and
            Mayorov, Nikolay and Nelson, Andrew R. J. and Jones, Eric and
            Kern, Robert and Larson, Eric and Carey, C J and
            Polat, {\.I}lhan and Feng, Yu and Moore, Eric W. and
            {VanderPlas}, Jake and Laxalde, Denis and Perktold, Josef and
            Cimrman, Robert and Henriksen, Ian and Quintero, E. A. and
            Harris, Charles R. and Archibald, Anne M. and
            Ribeiro, Ant{\^o}nio H. and Pedregosa, Fabian and
            {van Mulbregt}, Paul and {SciPy 1.0 Contributors}},
  title   = {{{SciPy} 1.0: Fundamental Algorithms for Scientific
            Computing in Python}},
  journal = {\nm{}},
  year    = {2020},
  volume  = {17},
  pages   = {261--272},
  doi     = {https://doi.org/10.1038/s41592-019-0686-2},
}

@article{h05,
    title = {Cuba - a library for multidimensional numerical integration},
    journal = {\cpm{}},
    volume = {168},
    number = {2},
    pages = {78-95},
    year = {2005},
    issn = {0010-4655},
    doi = {https://doi.org/10.1016/j.cpc.2005.01.010},
    author = {{Hahn}, Thomas R.}
}

@article{c08,
    title = {A one-parameter family of interpolating kernels for smoothed particle hydrodynamics studies},
    journal = {\jcp{}},
    volume = {227},
    number = {19},
    pages = {8523-8540},
    year = {2008},
    issn = {0021-9991},
    doi = {https://doi.org/10.1016/j.jcp.2008.06.014},
    author = {{Cabezón}, Rubén M. and {García-Senz}, Domingo and {Relaño}, Antonio}
}

@ARTICLE{s46,
       author = {{Sedov}, Leonid Ivanovich},
        title = "{Propagation of strong shock waves}",
      journal = {\jamm{}},
         year = 1946,
        month = jan,
       volume = {10},
        pages = {241-250}
}

@techreport{k00,
  author = {{Kamm}, James R.},
  institution = {\lanl{}},
  title = {Evaluation of the Sedov-von Neumann-Taylor Blast Wave Solution},
  year = 2000
}

@ARTICLE{v79,
       author = {{van Leer}, Bram},
        title = "{Towards the Ultimate Conservative Difference Scheme. V. A Second-Order Sequel to Godunov's Method}",
      journal = {\jcp},
         year = 1979,
        month = jul,
       volume = {32},
       number = {1},
        pages = {101-136},
          doi = {https://doi.org/10.1016/0021-9991(79)90145-1}
}

@ARTICLE{v82,
       author = {{van Albada}, G.~D. and {van Leer}, B. and {Roberts}, Jr., W.~W.},
        title = "{A comparative study of computational methods in cosmic gas dynamics}",
      journal = {\aaa{}},
         year = 1982,
        month = apr,
       volume = {108},
       number = {1},
        pages = {76-84}
}

@ARTICLE{h18,
       author = {{Hubber}, D.~A. and {Rosotti}, G.~P. and {Booth}, R.~A.},
        title = "{GANDALF - Graphical Astrophysics code for N-body Dynamics And Lagrangian Fluids}",
      journal = {\mnras{}},
         year = 2018,
        month = jan,
       volume = {473},
       number = {2},
        pages = {1603-1632},
          doi = {https://doi.org/10.1093/mnras/stx2405}
}






\end{document}